\documentclass[a4paper,11pt]{article}
\pdfoutput=1 

\usepackage{jinstpub} 

\usepackage{DRtubes-defs}
\usepackage[T1]{fontenc} 
\usepackage{xspace}
\usepackage{xcolor,colortbl}
\usepackage{hyperref}
\usepackage{subfigure}
\usepackage{tikz}
\usepackage{doi}
\usepackage{lineno}

\title{Exposing a fibre-based dual-readout calorimeter to a positron beam}

\author[a,b]{N. Ampilogov}
\author[a,b]{S. Cometti}
\author[c,d]{J. Agarwala}
\author[e]{V. Chmill}
\author[d]{R. Ferrari}
\author[d]{G. Gaudio}
\author[f]{P. Giacomelli}
\author[a,b]{A. Giaz}
\author[e]{A. Karadzhinova-Ferrer}
\author[g]{A. Loeschcke-Centeno}
\author[c,d]{A. Negri}
\author[h]{L. Pezzotti}
\author[d]{G. Polesello}
\author[a]{E. Proserpio}
\author[h]{A. Ribon}
\author[a,b]{R. Santoro}
\author[g]{I. Vivarelli}

\affiliation[a]{Universit\'a dell'Insubria}
\affiliation[b]{INFN Milano}
\affiliation[c]{Universit\'a degli Studi di Pavia}
\affiliation[d]{INFN Pavia}
\affiliation[e]{RBI Zagreb}
\affiliation[f]{Universit\'a degli Studi and INFN Bologna}
\affiliation[g]{University of Sussex}
\affiliation[h]{CERN}

\emailAdd{giacomo.polesello@cern.ch}
\emailAdd{i.vivarelli@sussex.ac.uk}

\abstract{A prototype of a dual-readout calorimeter using brass capillary tubes surrounding scintillating and clear plastic optical fibres was tested using beams of particles with energies between 10 and 100 GeV produced by the CERN SPS. The scope of the test was to characterise the performance of the tube-based detector response to positrons in terms of response linearity, energy resolution, and lateral granularity. After calibrating the detector and processing the output signal to correct for the energy dependency on the particle impact point, the linearity of the measurement was found to be better than 1\%. The response to positron was compared to that predicted by a Geant4-based simulation, finding good agreement both in terms of energy resolution and shower profile. These results confirm the validity of the tube-based mechanical option and SiPM readout as a promising one for future developments.
 }


\begin{document} 

 \maketitle

\flushbottom

\section{Introduction}
\label{sec:introduction}
One of the requirements of the desired physics programme of future $e^+e^-$ colliders (such as, for instance, \mbox{FCC-ee} and CEPC \cite{FCC:2018evy,CEPCStudyGroup:2018ghi}) is a high-precision measurement of the jet four-momenta. Dual-readout calorimetry relies on a dual sampling of the calorimeter signal, with two sensitive media having a different $h/e$ ratio. The combined information from the two signals allows for the effective correction of fluctuations of the hadronic shower's electromagnetic fraction, therefore dramatically improving the energy measurement while recovering the linearity of the response to hadrons. Dual-readout calorimetry~\cite{Lee:2017xss} is a well-established technique: a 20-year-long experimental programme established the proof-of-principle~\cite{Akchurin:2005eu,Akchurin:2005an,Akchurin:2005rs,Akchurin:2013yaa,Lee:2017shn, Akchurin:2014aoa},  converging on the solution of employing two sets of optical fibres embedded in the absorber, running nearly parallel to the direction of the incoming particles. Scintillating fibres are utilised to sample all charged components of the shower, while plastic undoped fibres collect Cherenkov light emitted mainly by electrons and positrons, and are therefore mostly sensitive to the electromagnetic component of the hadronic shower. Recent developments of the dual-readout technique witnessed the test of silicon photomultipliers (SiPMs) as light detectors capable of reading out individual fibres~\cite{Antonello:2018sna}. A simulation of a full 4$\pi$ dual-readout calorimeter, using a mechanics and geometry different from those tested in this paper is documented in Refs. \cite{lorenzoPhD, Pezzotti:2022ndj}, where 
a hadronic energy resolution of $\sigma/E \sim$30$\%$/$\sqrt{E}$ for single hadron and $\sigma/E\sim$38$\%$/$\sqrt{E}$ for jets was demonstrated, while maintaining good electron and photon energy resolution (of the order of $\sigma/E \sim 15\%/ \sqrt{E}$).  The baseline IDEA detector concept for future  $e^+e^-$ colliders~\cite{Gaudio:2022jve} utilises a dual-readout fibre-based calorimeter for the energy and position measurement of electrons, photons and hadrons. An option utilising a crystal-based dual-readout electromagnetic section in front of the fibre-based calorimeter was also studied, and its performance is described in Refs.~\cite{Lucchini:2020bac, Lucchini:2022vss}.

This paper focuses on the test of a new prototype employing a new scalable concept for the mechanical construction: the optical fibres are inserted into individual cylindrical brass capillary tubes, which are glued together to form calorimeter modules.The capillary tubes offer a flexible solution for large-scale construction at an affordable cost. A dual-readout calorimeter using brass capillary tubes as absorber and two types of optical fibres as active medium was built~\cite{Karadzhinova-Ferrer:2022paf} and tested using beams of particles during two different test-beam campaigns. The size of the prototype was such that electromagnetic showers can be efficiently contained. The objective of the test beams was to assess the calorimeter prototype performance in terms of linearity and resolution of its response to electrons. The excellent lateral granularity offered by the capillary tube solution is also exploited to perform a measurement of the electron shower shape. Section~\ref{sec:det_descr} describes the details of the experimental setup, including the calorimeter layout and its readout, and the setup of the auxiliary detectors used to select electrons among the beam particles. Section~\ref{sec:calibration} describes the calibration procedure followed to first equalise the response of the calorimeter modules, then calibrate the whole calorimeter to the electromagnetic scale. The calorimeter response to positrons is discussed in Section~\ref{sec:results}, and finally conclusions are drawn in Section~\ref{sec:conclusions}.

\section{Experimental setup and detector simulation}
\label{sec:det_descr}
\begin{figure}[htb]  
\begin{center}
    \subfigure[]{
        \includegraphics[width=0.35\textwidth]{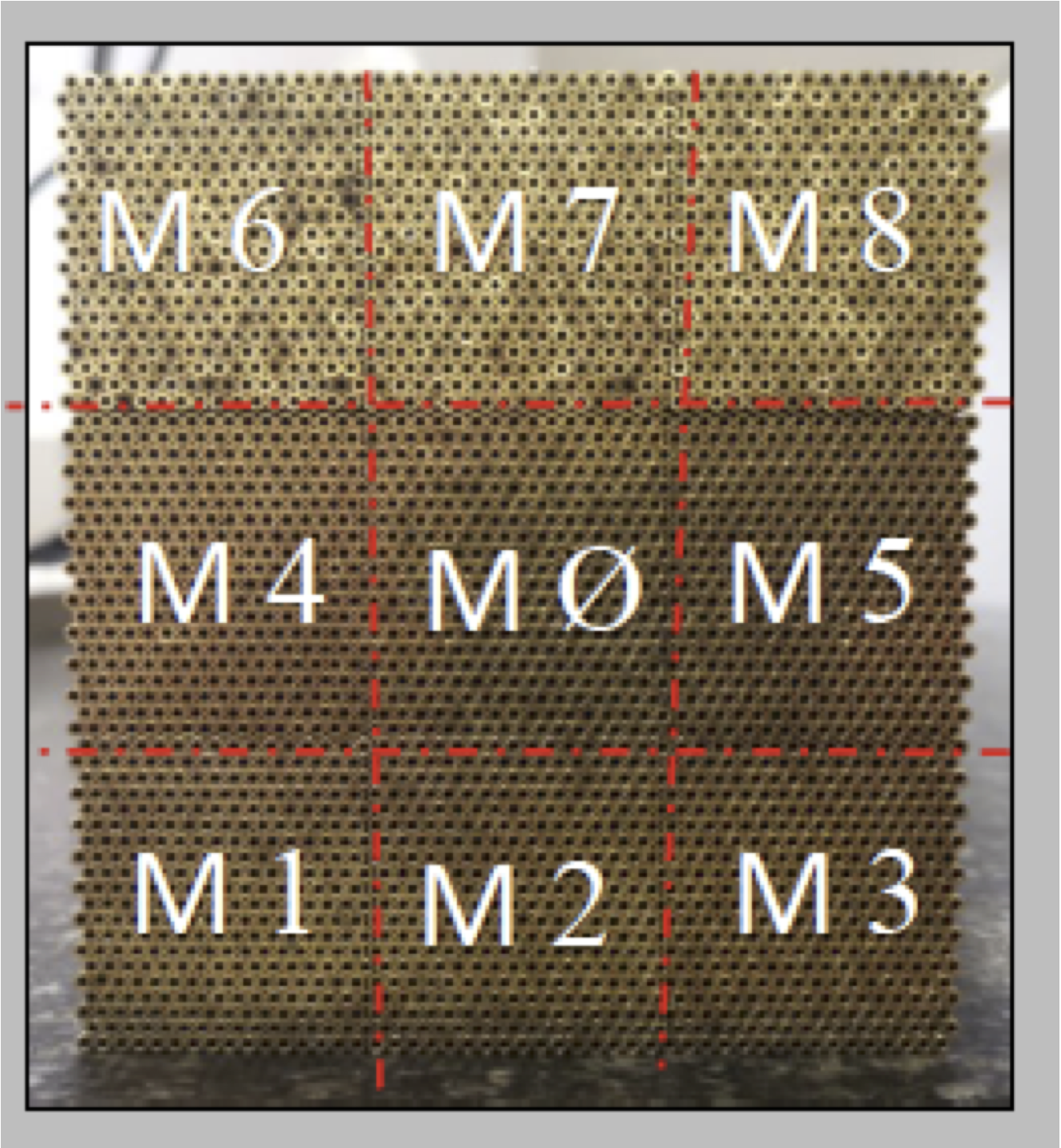}
    }
\hfill
     \subfigure[]{
        \includegraphics[width=0.45\textwidth]{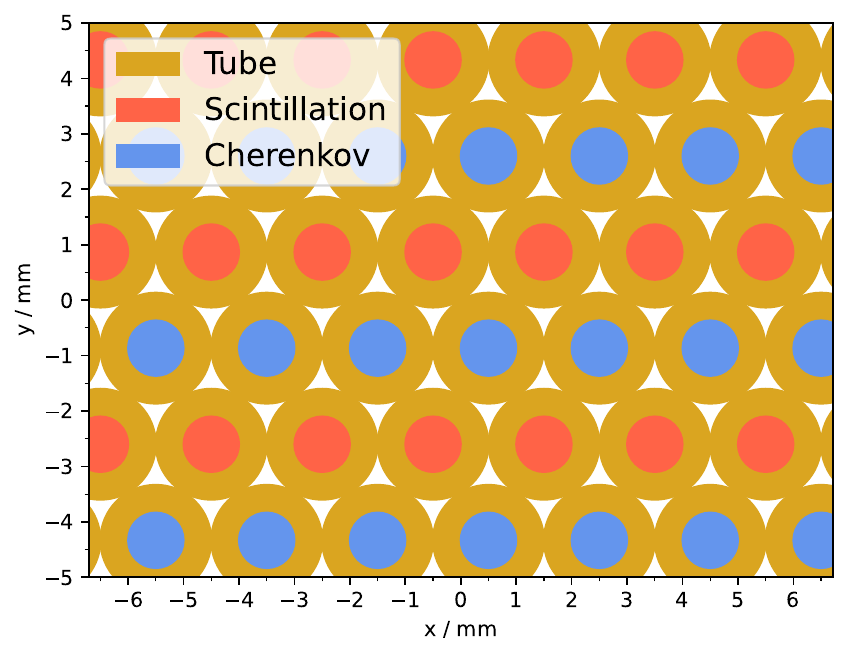}
     }
     \hfill
     \subfigure[]{
        \includegraphics[width=0.72\textwidth]{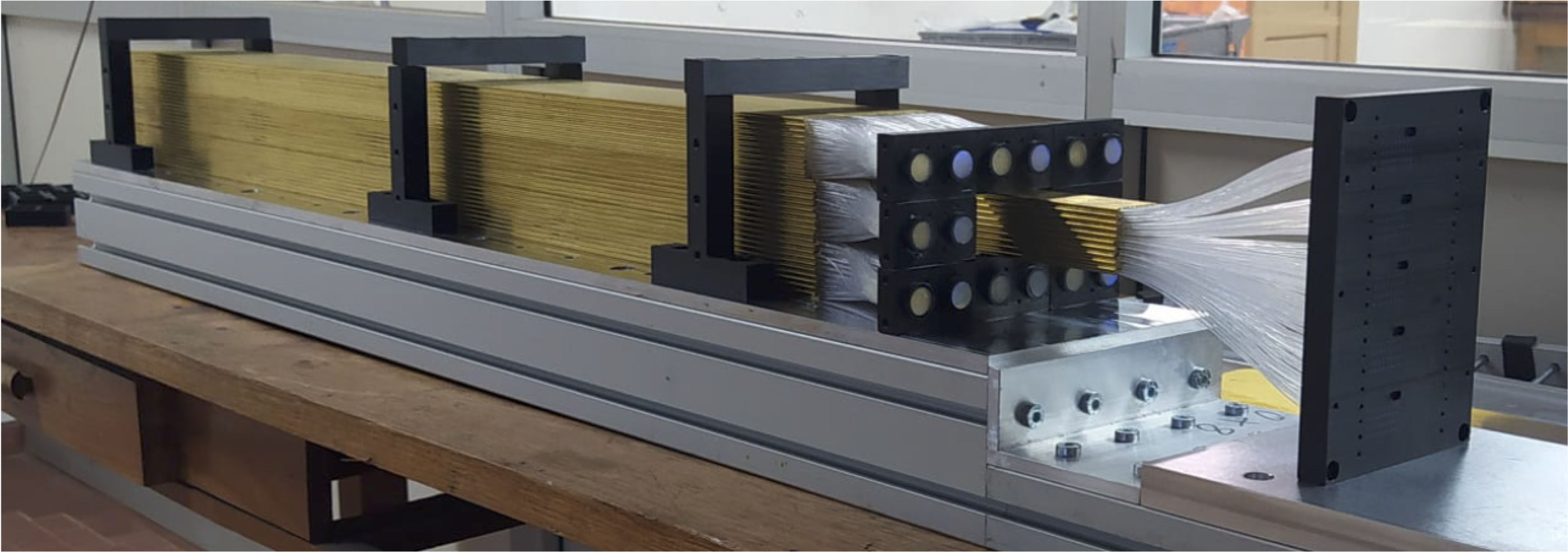}
     }
     \end{center}
\caption{(a) Front view of the prototype and (b) sketch of the front face of the calorimeter detailing the relative positions of the Cherenkov and scintillating fibres. (c) Lateral view of the prototype showing how the fibres from the external modules are bundled to match the PMTs' window while the longer fibres from the central module are connected to a patch panel to be interfaced with SiPMs.}
\label{fig:prototype}
\end{figure}

The prototype tested on beam in 2021 is shown in Figure~\ref{fig:prototype}. Its size is about $100 \times 10 \times 10\ \mathrm{cm}^3$. It consists of nine identical modules, arranged as detailed in Figure~\ref{fig:prototype}(a). Each module is made of 320 brass (63\% Cu, 37\% Zn) capillaries (with external diameter of 2 mm and internal diameter of 1.1 mm)  equipped, alternatively, with scintillating and clear undoped optical fibres. Both sets of fibres have an external diameter of 1 mm. The scintillating fibres are BCF-10 produced by Saint-Gobain. They have a polystyrene-based core and a single PMMA clad. The emission peak is at 432 nm, and the light yield is about 8000 photons per MeV. More information can be found at Ref.~\cite{BC10}. The clear undoped fibres are SK-40 from Mitsubishi. They have a PMMA resin core and a fluorinated polymer clad, and a numerical aperture of 0.5. More information can be found at Ref.~\cite{SK40}.

The brass absorber of the capillary tubes, accounts for about 66\% of the volume, while each type of fibre accounts for about 11\%. The remaining 12\% is occupied by air and glue. The Moli\`ere radius of such prototype is calculated to be 23.8~mm while the effective radiation length is 22.7~mm. The layout of the scintillating and clear fibres is shown in Figure~\ref{fig:prototype}(b). 

\begin{figure}[htb]
\begin{center}
    \subfigure[]{
\includegraphics[width=0.31\textwidth]{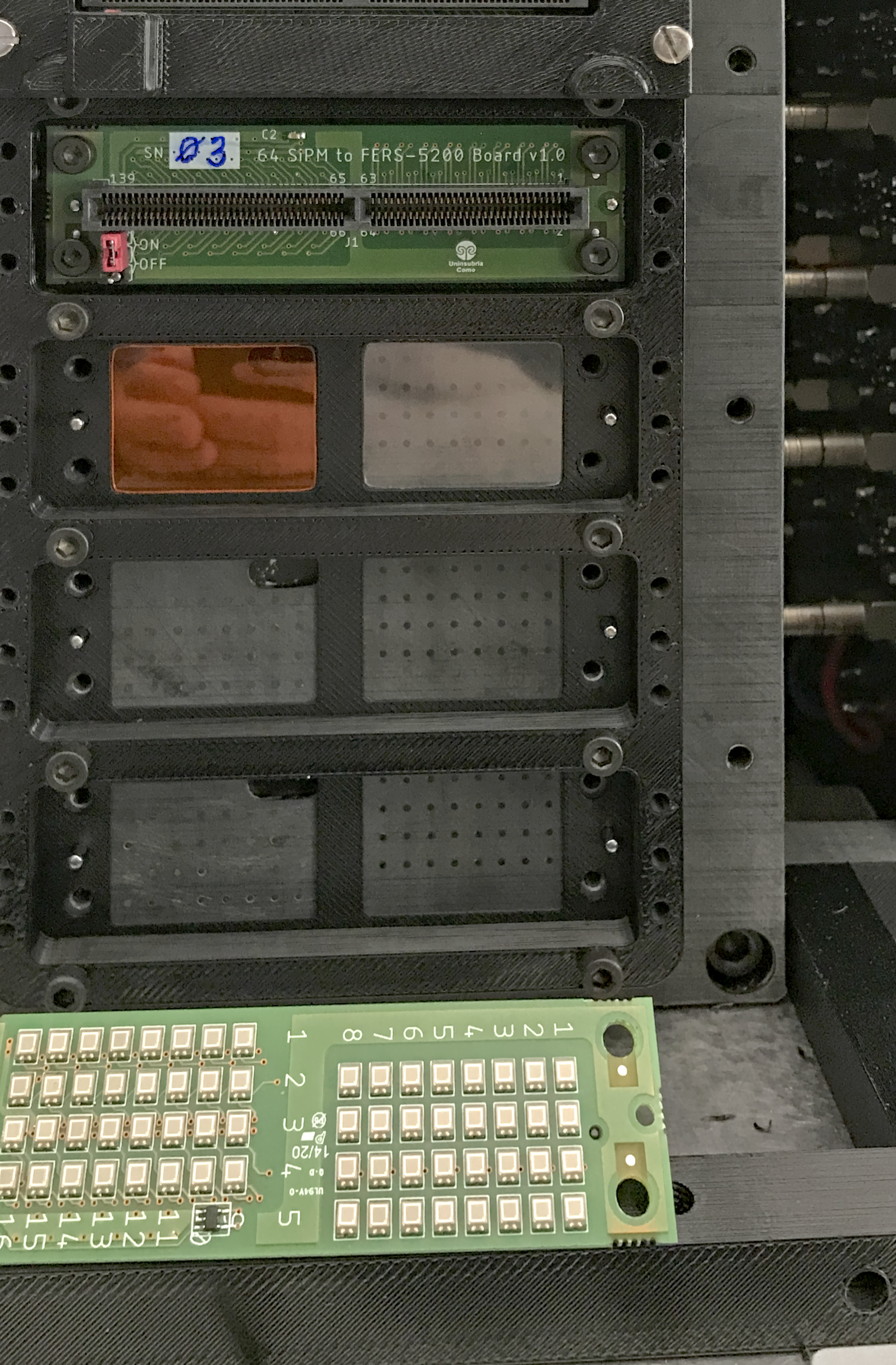}
    }
\hfill
     \subfigure[]{
     \includegraphics[width=0.65\textwidth]{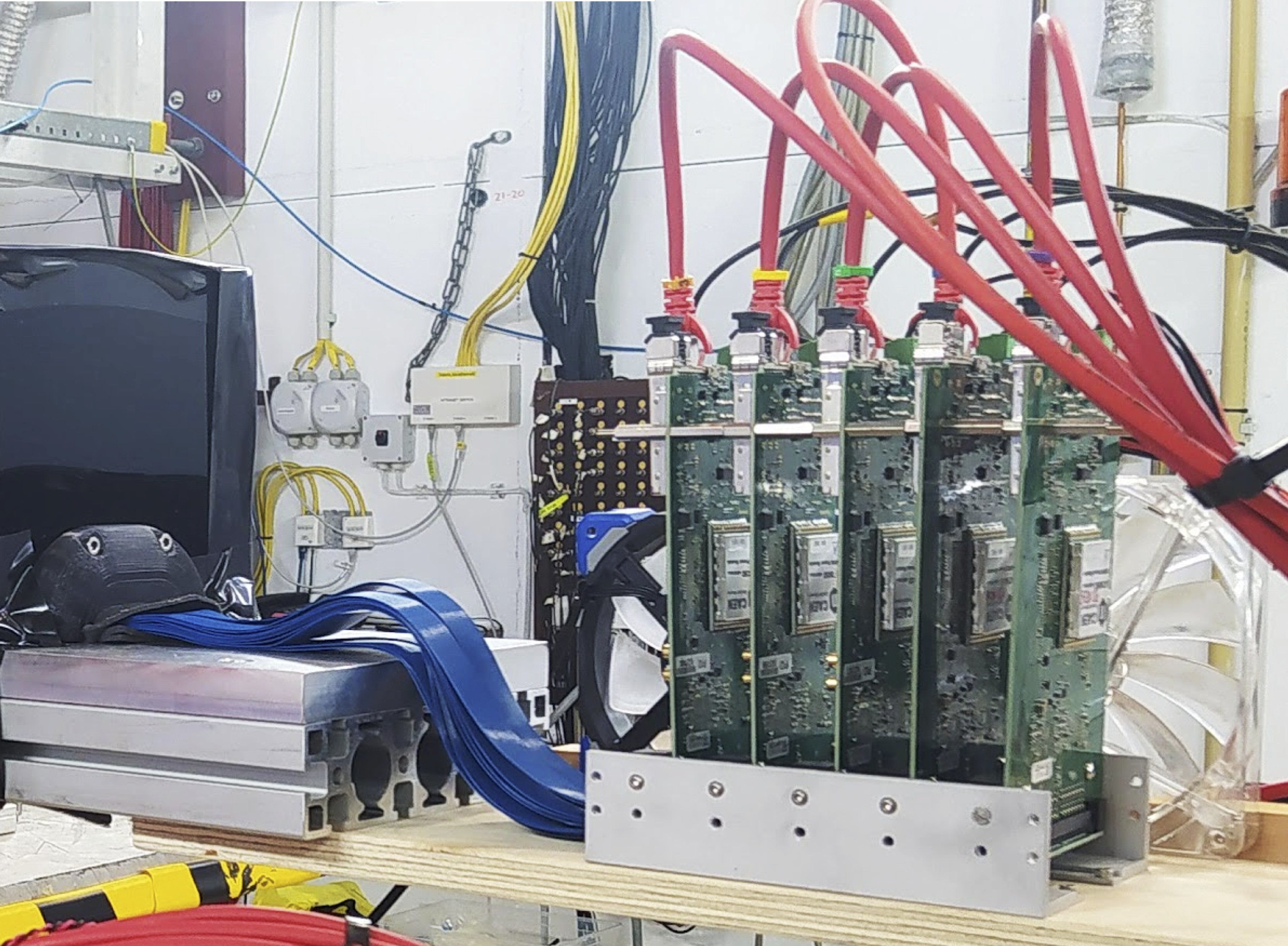}
     }
     \end{center}
\caption{(a) Back side of $\M{0}$ and a front-end board with 64 SiPMs before the installation. (b) The full system connected to the five~A5202~boards utilised to read out the 320 SiPMs.}
\label{fig:readout}
\end{figure}

The external modules $\M{1}-\M{8}$ are instrumented with Hamamatsu R8900 PMTs~\cite{R8900}. The scintillating and clear fibres are separated and bundled in two groups on the back side of each module to match the PMTs' window. A yellow filter (Kodak Wratten 3, with nominal transmission of about 7\% at 425~nm and 90\% at 550~nm, visible in Figure~\ref{fig:readout}(a)) is placed between the scintillating fibres and the detector to attenuate the scintillation signal and to cut off short wavelength components of the light: this helps reducing the calorimeter response dependence on the shower depth and starting point by selecting wavelengths with a longer fibre attenuation length. The PMTs are read out with V792AC QDC modules produced by CAEN S.p.A..

For the  central module $\M{0}$, each individual fibre is read out by a Hamamatsu SiPM (S14160-1315 PS~\cite{S14160}) with a sensitive area of 1.3 $\times$ 1.3 mm$^2$. Since almost 10$\%$ of the entire energy is released within one millimeter from the core of the shower (1--2 fibres) \cite{Antonello:2018sna,lorenzoPhD}, SiPMs with a wide dynamic range (i.e., 7284 cells, 15~$\mu$m pitch) were selected. SiPMs with compact packaging were not available at the time of the construction, therefore it was decided to fan out the fibres on the back sides of the calorimeter to match the front-end boards housing 64 SiPMs. The SiPMs on the front-end board are separated in two groups (32 SiPMs each) insulated with a light tight frame (Figure~\ref{fig:readout}(a)) to avoid optical crosstalk between the Cherenkov and scintillation signals.
As for the modules $\M{1}-\M{8}$, yellow filters are placed between the scintillating fibres and the SiPMs. In addition, for $\M{0}$ a transparent paper is used between the clear fibres and the SiPMs for mechanical reason and to avoid any air gap between the fibres and the light sensors.

The SiPM readout is based on the FERS system produced by CAEN S.p.A.~\cite{FERS:CAEN} that fully exploits the Citiroc~1A~\cite{CITIROC} performance: wide dynamic range, linearity and single photon detection ability even with SiPMs with small pitch size and small gain (1--3 $\times$ 10$^{5}$ at nominal settings). The SiPM readout setup is shown in Figure~\ref{fig:readout} (b). Each readout board (A5202) is equipped with two Citiroc~1A to operate 64 SiPMs. The signal produced by each SiPM feeds two~charge~amplifiers with tunable gains. The range accessible by one of the two amplifiers (namely the High Gain, HG in the following) is almost 10 times higher than the other (the Low Gain, LG). This feature allows two spectra for each SiPM to be stored on disk. The spectrum acquired via the HG chain is useful to analyse the multiphoton signal and to extract the ADC-to-photoelectrons (\phe) conversion constant. The LG spectrum extends the overall dynamic range. The settings for the two charge amplifiers were chosen to guarantee good quality HG spectra and a wide dynamic range, while maintaining an overlap between the signals acquired with the two different gains to be used for their intercalibration. The settings chosen allow signals from 1 to almost 4000 \phe to be read out. This corresponds to  about 55\% of the SiPM occupancy considering the microcells available in the sensitive area.

The prototype was qualified on beam at the Deutches Elektronen-SYnchrotron (DESY)~\cite{Diener:2018qap} using electron beams with energies $1\ \mathrm{GeV} \le \Ebeam \le 6\ \mathrm{GeV}$, and at CERN, using positron beams of the SPS H8 beam line with energies $10\ \mathrm{GeV} \le \Ebeam \le 100 \ \mathrm{GeV}$. The results of this paper use only data collected at CERN. 

\subsection{Beam setup}

A set of auxiliary detectors present on the beam line were used to improve particle identification and support the data taking. Figure~\ref{fig:beam_sketch} sketches the beam setup. 

\begin{itemize} 

\item Upstream the beam, two Cherenkov threshold counters~\cite{Dannheim:2013iea} were available. They were filled with $\mathrm{He}$. The pressure of the gas in the counters was chosen depending on the beam energy to optimise the separation between electrons and pions. 

\item Two scintillation counters ($\mathrm{T_1}$ and $\mathrm{T_2}$ in the following), each 2.5~mm thick, with an area of overlap of about  $4 \times 4\ \mathrm{cm^2}$ were used in coincidence ($\mathrm{T_1} \land \mathrm{T_2}$). A third scintillator counter ($\mathrm{T_3}$), placed downstream the beam, had a 10 mm radius hole. Its purpose was to veto particles travelling far from the beam axis. The particle trigger signal was defined to be $\left(\mathrm{T_1} \land \mathrm{T_2}\right) \land \bar{T}_3$ (physics trigger in the following). 

\item A pair of Delay Wire Chambers (DWCs) were used to be able to determine the location of the impact point of the beam particles at the calorimeter surface with a precision of a few mm, depending on the beam energy. DWC1 and DWC2 were placed respectively upstream and downstream the beam with respect to the trigger scintillators. 

\item A preshower detector (PS in the following), consisting of 5~mm of lead and a scintillator slab read out with a photomultiplier, was located at 285~cm from the face of the calorimeter. Its purpose was to help with the identification of high-energy positrons, which would be the only particles in the beam with a significant radiation probability in the lead. The positioning of the calorimeter at such a distance from the PS was forced due to access limitations to the available test-beam area.   

\begin{figure}[htb]
\begin{center}
\includegraphics[width=1.05\textwidth]{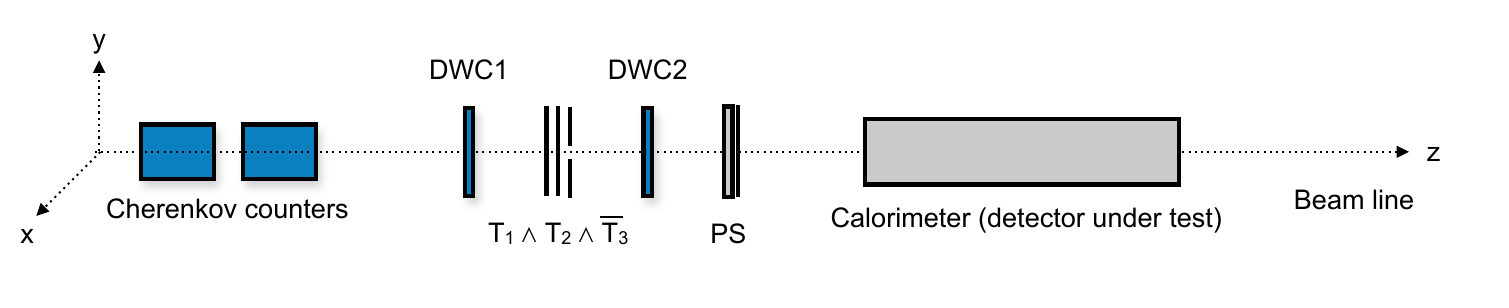}
     \end{center}
\caption{Sketch of the beam line setup. The diagram is not in scale.}
\label{fig:beam_sketch}
\end{figure}  


\end{itemize}

The signals from the auxiliary detectors and the PMTs reading modules $\M{1}-\M{8}$ were collected using a single data acquisition system. The system was set up so that every ten physics triggers, a ``pedestal'' trigger was produced. All trigger signals (physics and pedestals) were also sent to a second independent data acquisition system, reading out the signals from the SiPMs of the $\M{0}$ module. The synchronisation of the two data acquisition systems was done offline, by making use of the pedestal events.  

In the following, a right-handed orthogonal system of coordinates with the $z$-axis along the beam line, and with the $y$-axis pointing upwards is used. The origin of the coordinate system is taken to be on the front face of the calorimeter, at the geometrical centre of $\M{0}$.  The calorimeter prototype was placed on beam so that its longest side was forming an angle of about $1^{\circ}$ with the $z$-axis in the $x-z$ plane. This is to avoid channeling effects (particles entering and travelling long distances in an optical fibre, causing abnormally large scintillation signals). No angle was introduced between the calorimeter long axis and the $z$-axis in the $y-z$ plane. 

\subsection{Test beam simulation}

The full test beam setup (including the calorimeter prototype and the auxiliary detectors) was simulated using \Geant \verb|10.7.p01|~\cite{Agostinelli:2002hh}. The software package can simulate the detector response in a ``fast'' or ``full'' mode. For the full mode, the photon propagation within the optical fibres is completely delegated to \Geant. In the fast mode, such propagation is parametrised. In either mode, the scintillation and Cherenkov light yields have been tuned to correctly reproduce the average amount of light. To take into account the Poissonian fluctuations of light emission, the simulated light emission at every hit was smeared accordingly. The simulation results in the following sections have been obtained using the fast simulation mode, using the \verb|FTFP_BERT| physics list. The response of the scintillating fibres takes into account the Birks' effects on light emission. In particular, the description of polystyrene-based scintillating fibres incorporates a Birks constant value of 0.126 mm/MeV. Nevertheless, the inclusion of the Birks Law correction into the Monte Carlo description primarily impacts the outcomes related to hadron-induced particle showers. This is due to the fact that massive charged particles within such showers may experience exceptionally high energy deposition rates ($dE/dx$) compared to the lighter charged particles present in showers induced by electrons, positrons, or gamma rays. Consequently, the inclusion of the Birks Law correction in the simulation yields small differences in the subsequent results.
For both signals, the output of the simulation is the number of \phe expected at the end of each optical fibre. No emulation of the SiPMs is included - therefore the simulation does not include, e.g., the noise introduced by the readout electronics.

\section{Calorimeter response equalisation and calibration}
\label{sec:calibration}

The calibration of the prototype was performed in three steps. First, the gain of all SiPMs in $\M{0}$ was equalised, and the conversion factor between ADC counts and \phe derived, by making use of the SiPM multiphoton spectrum. Then, the response of all modules $\M{0}$-$\M{8}$ was equalised by making use of a positron beam with an energy of 20 GeV. Finally, the overall calorimeter energy scale was set by looking at the response of the whole calorimeter prototype to beams of positrons.

\subsection{SiPM equalisation using the multiphoton spectrum}

A first equalisation of the SiPMs was performed in the laboratory before the beam tests, by making use of an ultra  fast LED emitting at 420 nm. The SiPM response was equalised by applying the same overvoltage and by tuning the amplifier settings. The procedure allowed to operate all SiPMs with about the same photon detection efficiency (PDE) and with signals equalised in amplitude. The SiPMs were operated with a voltage set at $+7\ \mathrm{V}$ over the breakdown voltage. Even if it is not a typical setting for a SiPM, it guarantees a stable PDE (against small temperature variations) and a multiplication factor in the avalanche region of the order of $0.5 \cdot 10^{6}$ per each detected photon. The spurious effects (i.e. dark count rate and crosstalk) have limited impact on the measurements. 

\begin{figure}[ht]
\begin{center}
    \subfigure[]{
        \includegraphics[width=0.45\textwidth]{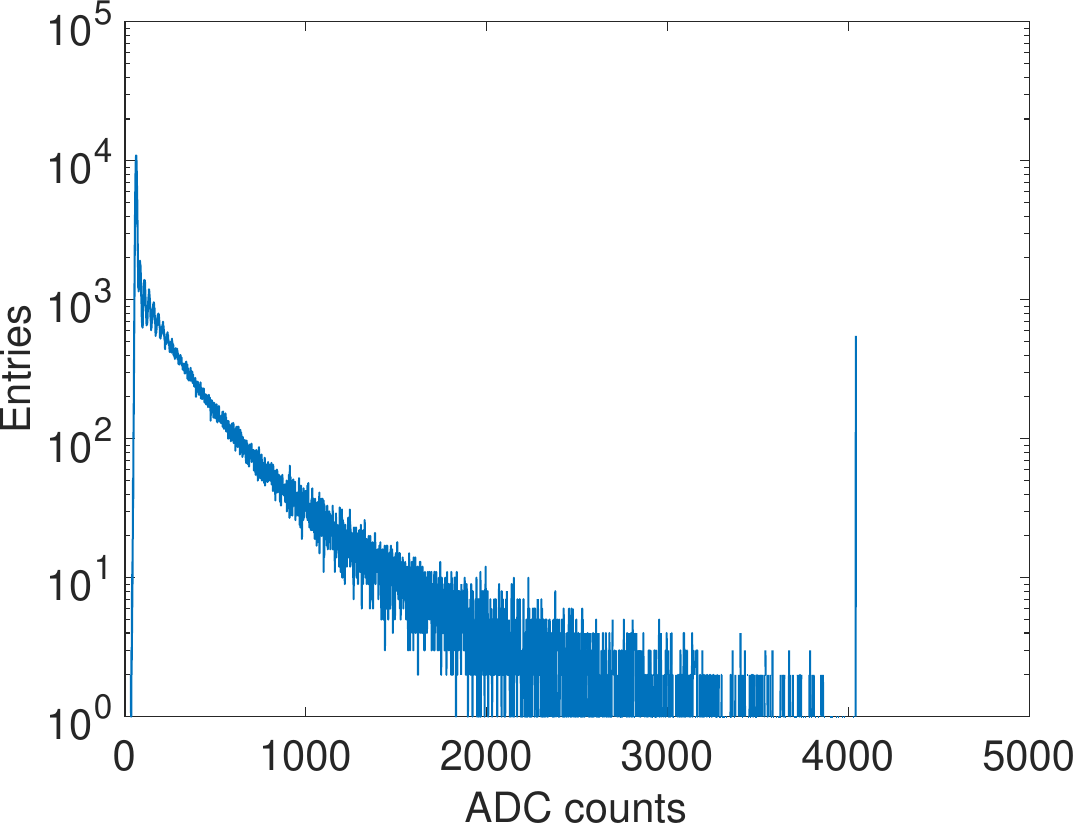}
    }
\hfill
     \subfigure[]{
        \includegraphics[width=0.45\textwidth]{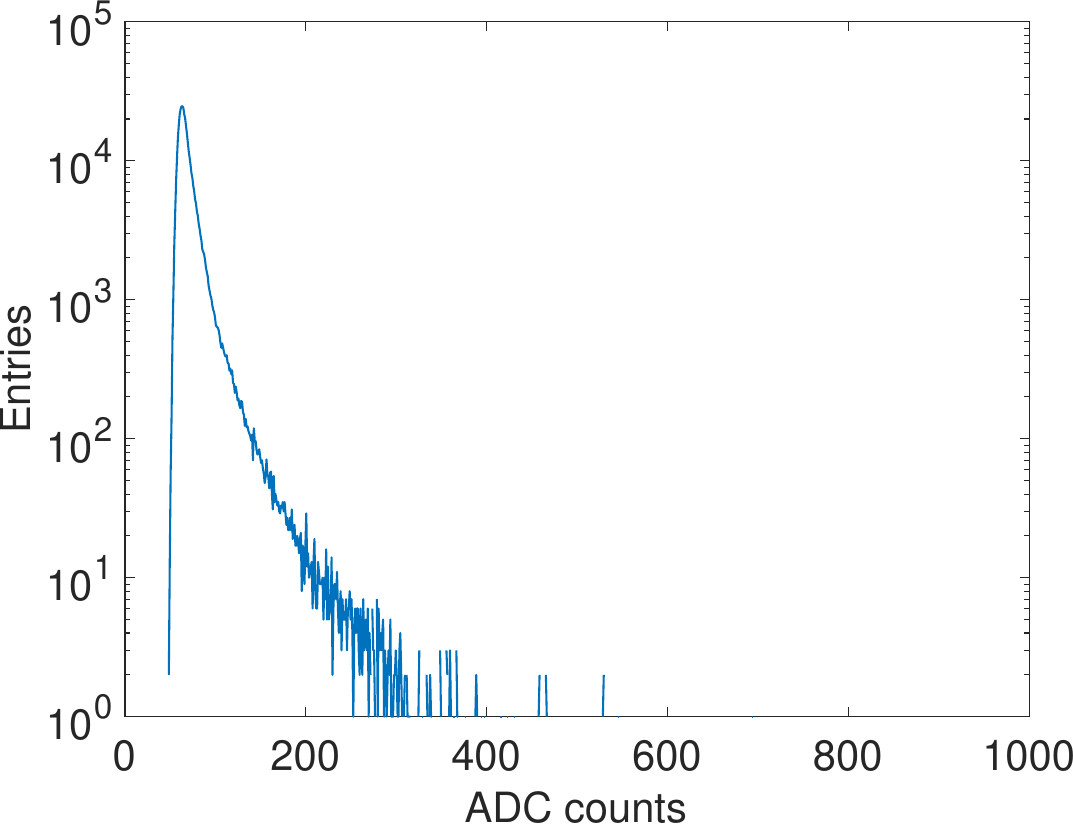}
     }
\caption{(a) High gain (HG) and (b) low gain (LG) spectra obtained for the same SiPM with 40-GeV electrons.}
\label{fig_spectrum}
\end{center}
\end{figure}  

\begin{figure}[ht]
\begin{center}
\subfigure[]{
    \includegraphics[width=0.44\textwidth]{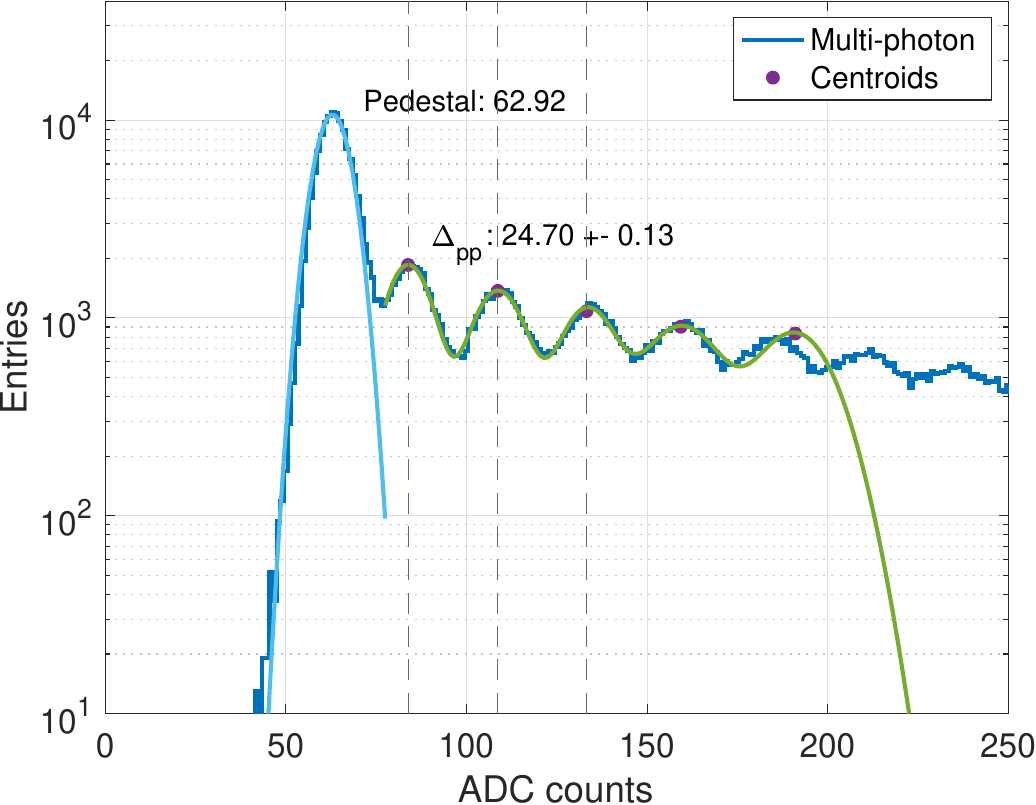}
    } \hfill
\subfigure[]{
    \includegraphics[width=0.50\textwidth]{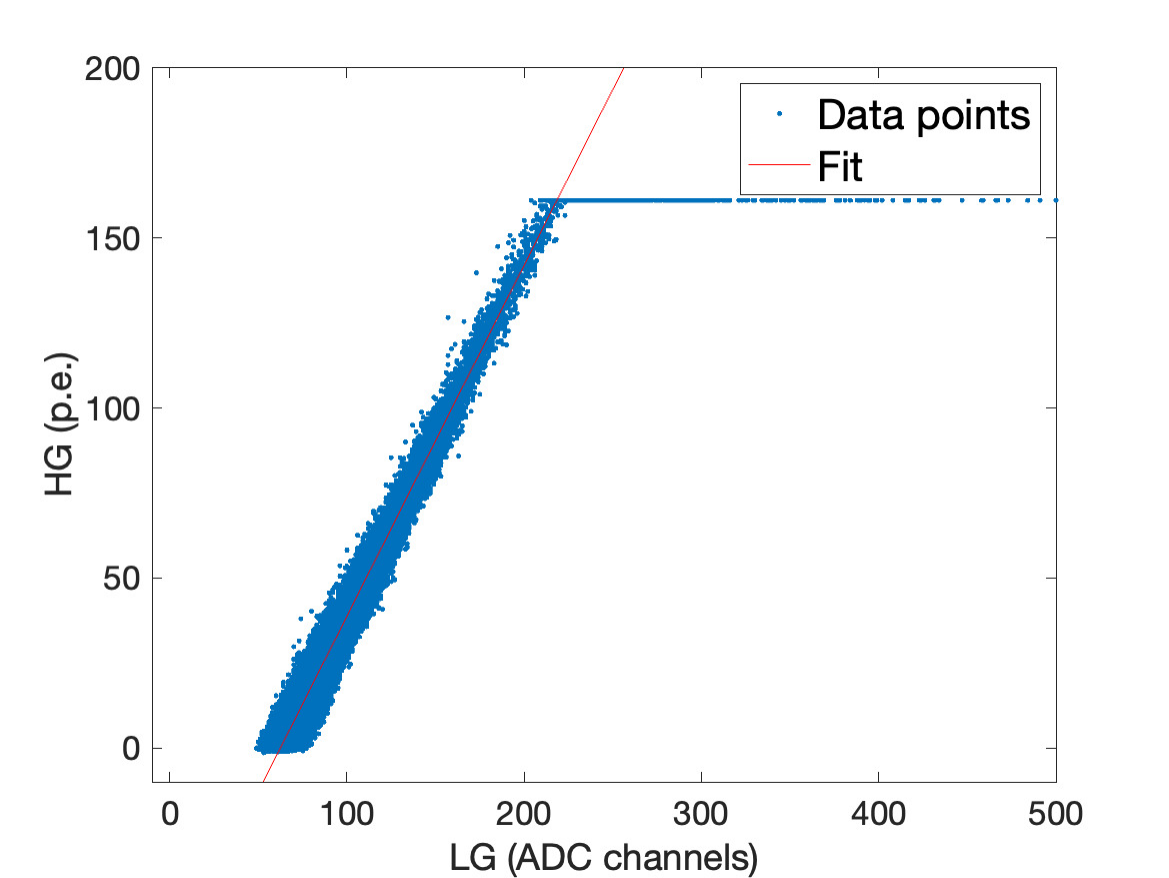}
    }
\caption{(a) Zooming of the region below 250 ADC counts of the HG spectrum in \Figure\ref{fig_spectrum}(a) showing the multiphoton fit. (b)~Scatter plot of HG (in \phe) signal against the LG signal for the same SiPM.}
\label{fig_calibration}
\end{center}
\end{figure}

Calibration in \phe of the LG response is required in order to sum signals from different SiPMs and to correct for non-linearity due to the limited number of cells available in each detector, if needed. 
Figure~\ref{fig_spectrum} shows the typical spectra measured in the two gains by one SiPM connected to a scintillating fibre in response to positrons with an energy of 40 GeV.
The pedestal, the multiphoton spectrum and the ADC saturation are clearly visible for the HG plot. The saturation is not affecting the measurement since the information is still available in the LG spectrum. The strategy used to calibrate both spectra in \phe was the following: 

\begin{itemize}
\item	The pedestal and the multiphoton distribution were fitted with Gaussian functions. The results were used to convert the ADC channels in p.e. by using the mean value of the pedestal and the average peak-to-peak separation obtained by fitting three consecutive peaks (Figure~\ref{fig_calibration}(a)).
\item	The HG values, converted in \phe, were plotted against those of the ADC counts of the LG channel (\Figure~\ref{fig_calibration}(b)). The points in the plot exceeding 125 \phe were not considered in the fit and the slope was used to extract the ADC-to-\phe conversion for the LG even if the multiphoton structure was not accessible in this regime. 
The typical conversion factor was about 1 \phe/ADC count with uncertainties of the order of 0.1\%. 
\end{itemize}

The procedure was performed for all SiPMs and the parameters were extracted on a run-by-run basis, without having the need to collect dedicated data for calibration. The stability of the SiPM calibration performed with the multiphoton spectrum was monitored over long periods of time: variations below 2$\%$ on the pedestal values and on the peak-to-peak separation  were measured for all SiPMs. It was therefore decided to use a single calibration constant for each SiPM for the full data taking period. The signal used in the following for the data analysis is the SiPM signal in \phe from the HG, unless the HG is found to be saturated, in which case the LG signal in \phe is used. 


\subsection{Calorimeter response equalisation}

The second step in the calibration was the equalisation of the modules $\M{0}-\M{8}$. The equalisation was performed by means of nine different runs, each collected by translating the calorimeter prototype so that a beam with energy $\Ebeam = 20\ \mathrm{GeV}$ was hitting in the centre of a different module. The positron component of the beam was selected by making use of the auxiliary Cherenkov counters. Selections are applied to both counters to reject events with a signal compatible with the pedestal in each of them. The beam composition could be studied by making use of the counters and of the PS auxiliary detector: it was estimated that the Cherenkov selection yields a nearly pure positron beam with a hadron contamination below 5\% for $\Ebeam = 20\ \mathrm{GeV}$.

The equalisation procedure assumed an equal tower response to positrons, in runs where the beam is hitting the module centre. The individual module containment was estimated (from the simulation) to be $\epsilon_{\mathrm{module}} = 72\%$. We define the average scintillation (Cherenkov) response in ADC counts of a module $i$ ($i = 1,8$) to a positron beam of energy $\Ebeam$ hitting its centre as $A^{S}_{\M{i}} \left(\Ebeam\right)$ ($A^{C}_{\M{i}}\left(\Ebeam\right)$). The response of the module in \phe $P^{S,C}_{\M{i}} \left(\Ebeam\right)$ is obtained by multiplying $A^{S,C}_{\M{i}}$ by a constant $a_i^{S,C}$ independent of the beam energy. The constants $a_i^{S,C}$ are determined by imposing that 

\begin{align*}
    P^{S,C}_{\M{i}} \left(20\ \mathrm{GeV}\right) = a_i^{S,C} \times A^{S,C}_{\M{i}}\ \left(20\ \mathrm{GeV}\right) = P^{S,C}_{\M{0}} \left(20\ \mathrm{GeV}\right),
\end{align*}

\noindent that is, by imposing that the module response is the same as that of module $\M{0}$ in identical conditions.  

\begin{figure}[ht]
\begin{center}
\centering
\includegraphics[width=1.\textwidth]{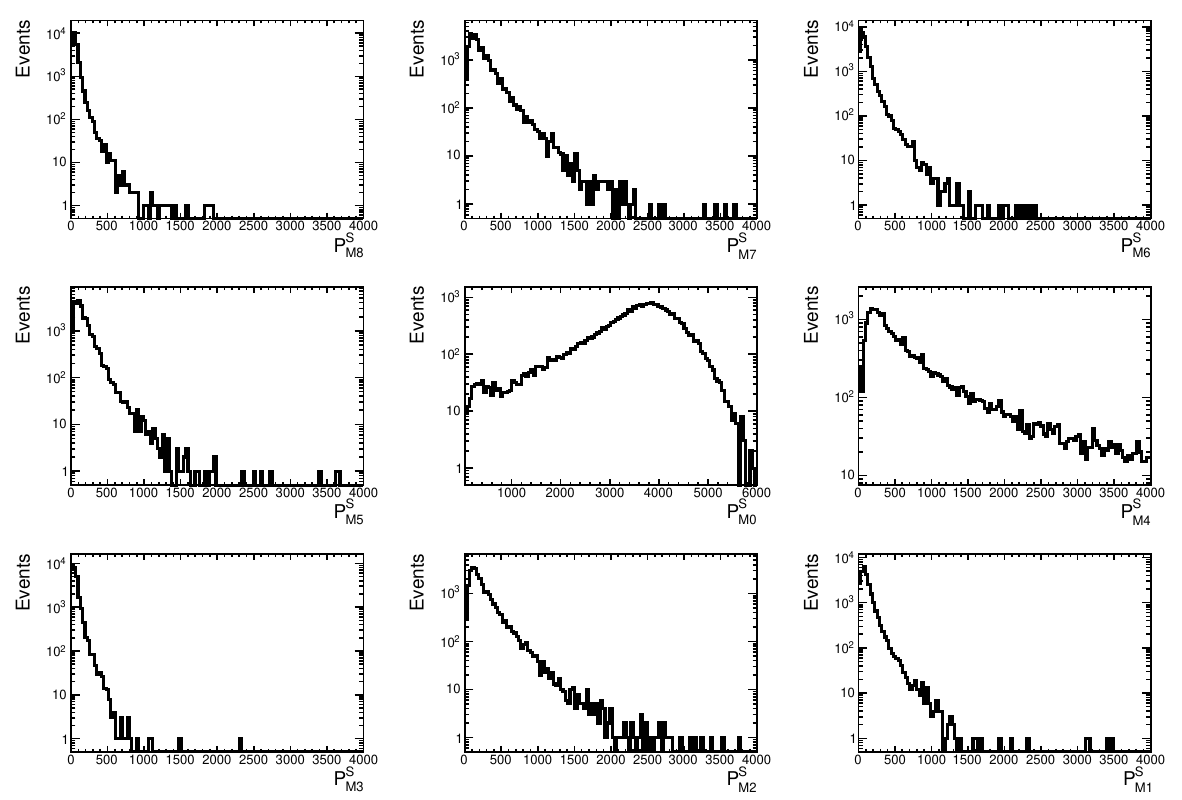}
\caption{Energy deposited in each module during a run where the 20-GeV positron beam was aimed at the centre of $\M{0}$ for the scintillation channel. The scale on the $x$-axis corresponds to \phe, while for the modules $\M{1}-\M{8}$ the response is equalised to that of $\M{0}$ following the procedure described in the text.} 
\label{fig:calo_eq_response_S}
\end{center}
\end{figure}

\begin{figure}[ht]
\begin{center}
\centering
\includegraphics[width=1.\textwidth]{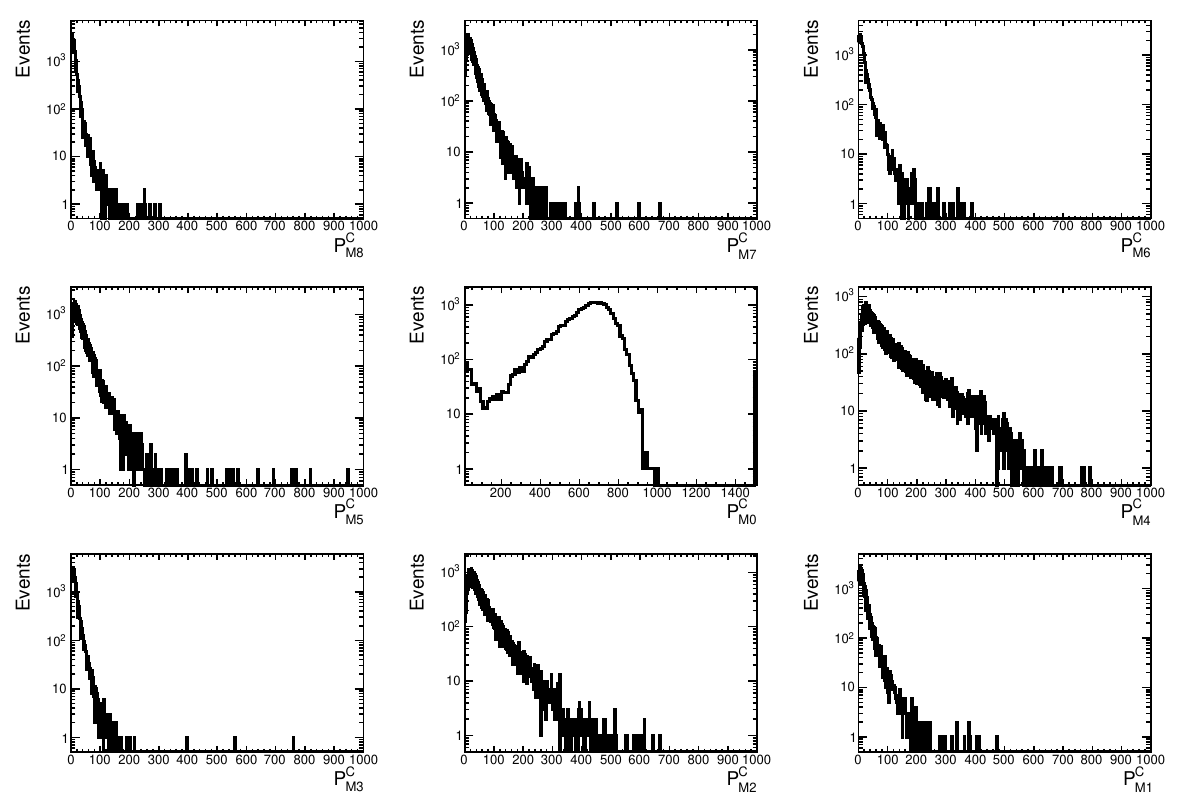}
\caption{Same as figure~\ref{fig:calo_eq_response_S}, but for the Cherenkov channel.} 
\label{fig:calo_eq_response_C}
\end{center}
\end{figure}

Figures~\ref{fig:calo_eq_response_S} and \ref{fig:calo_eq_response_C} show the scintillation  and Cherenkov signals in each tower of the calorimeter after equalisation, when the 20-GeV positron beam is hitting in $\M{0}$. The left-right asymmetry in the energy deposit is due to the angle between the beam line and the prototype main axis.  

After the equalisation was performed, all modules were expected to respond the same to a signal of a positron beam, and the signal was referred to the numbers of \phe seen by $\M{0}$ when hit by a positron beam of the same energy.  It is interesting to evaluate how many \phe $\M{0}$ sees per GeV of input energy. The values of the \phe to GeV conversion factor for $\M{0}$, $F^{S(C)}_{\mathrm{PheToGeV}}$ are extracted as 

\begin{align*}
F^{S,C}_{\mathrm{PheToGeV}} = \frac{P^{S,C}_{\M{0}} \left(20\ \mathrm{GeV}\right)}{\epsilon_{\mathrm{module}} \times 20\ \mathrm{GeV}}
\end{align*}

The measured values are $F^{S}_{\mathrm{PheToGeV}} = 277$, $F^{C}_{\mathrm{PheToGeV}} = 48$. These values are found to be compatible with the ones obtained in Ref.~\cite{Giaz:2023qsm}, which were obtained as an average over many runs from both test beam periods. 

\subsection{Calorimeter calibration}

The final step for the prototype calibration was to rescale the response of each module by a single constant ($\delta_{S}$ for the scintillation signal and $\delta_{C}$ for the Cherenkov signal), so that the total energy measured in the calorimeter corresponded to the beam energy separately for the scintillation and Cherenkov signal\footnote{The simulation predicts a shower containment of the full prototype of $\epsilon = 94\%$ at $\Ebeam = 20 \ \mathrm{GeV}$. This number was found to be nearly independent with $\Ebeam$. This lateral leakage was therefore de-facto reabsorbed in the determination of $\delta_{S,C}$.}. The constants $\delta_{S}$ and $\delta_{C}$ are determined as 

\begin{align*}
    \delta_{S,C} = \frac{20\ \mathrm{GeV}}{\langle\sum_{i = 0}^{8} P^{S,C}_{\M{i}}\rangle},
\end{align*}

\noindent where the average is computed over all selected positrons from a 20-GeV run with the beam pointing to the geometric centre of $\M{0}$.

The individual module energy (in GeV) can therefore be defined as 

\begin{align*}
    E^{S,C}_i = \delta_{S,C}\times P^{S,C}_{\M{i}},
\end{align*}







\section{Results}
\label{sec:results}
After the calibration procedure was applied, the data were analysed to measure the performance of the 
prototype for positrons. These were characterised in terms of the linearity of the response and its resolution  as a function of the beam energy. Moreover, exploiting the high granularity provided by the SiPM readout, the shape of the electromagnetic showers in the calorimeter was measured and compared to the \Geant simulation.

\subsection{Positron selection}
\label{sec:elec_selection}

Despite a beam configuration optimised to yield positrons, the beam of the H8 beam line at CERN contained in general positrons, muons and hadrons, in proportions which are strongly energy dependent. The definition of a procedure to select positrons was necessary. The first step was the definition of a fiducial area for the impact of the beam on the calorimeter ensuring good containment of the beam in $\M{0}$.

Due to a malfunctioning, the DWCs were determined to have a space resolution of about 2~mm, not adequate for studies of the dependence of the calorimeter response on the impact point and angle of the beam. The chambers were thus only used to clean the beam spot by requiring that the radius of the beam in both chambers was smaller
than 15~mm, and that for both coordinates, the difference of
the coordinates measured in the two chambers was less than $3\ \mathrm{mm}$.

Studies which require a precise knowledge of the position of the
shower inside the calorimeter are based on the barycentre of the 
shower (\Xcalo,\Ycalo). These variables are defined as:
\begin{equation}
\Xcalo=\frac{\sum_i x_i E_i}{\sum_i E_i}; \hspace{1cm} 
\Ycalo=\frac{\sum_i y_i E_i}{\sum_i E_i}
\label{formula1}
\end{equation}

\noindent where the sum runs on all of the 320 fibres in $\M{0}$,
$E_i$ is the energy deposited in each fibre, with coordinates
($x_i$, $y_i$). In order to guarantee good containment in the central module, an event-by-event selection was applied to the position of the barycentre. 

\begin{figure}[htb]
\begin{center}
        \includegraphics[width=0.5\textwidth]{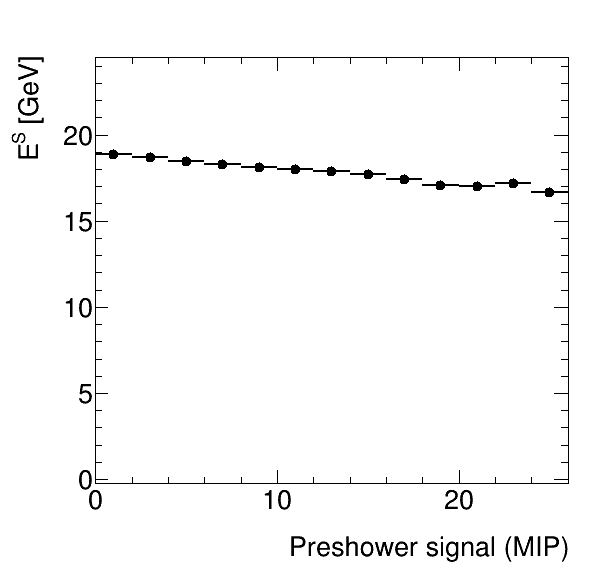}
\caption{Average value of the measured energy in the calorimeter for the scintillation signal as a function of the size of the signal in the PS detector.}
\label{fig:psene}
\end{center}
\end{figure}

For the selection of a pure positron beam, two Cherenkov counters and a PS detector were available, as described in Section~\ref{sec:det_descr}. A fraction of the positrons starts showering in the PS,
yielding a signal proportional to the number of secondary electrons and positrons produced in the shower.
Positrons can therefore be separated from more massive particles (hadrons and muons) by requiring a PS signal larger than that of a Minimum Ionising Particle (MIP).
However, the secondary electrons/positrons and photons emerge 
from the PS at an angle with respect 
to the primary positron. Given the significant distance of the
PS from the calorimeter face, the fraction of the beam energy 
collected in the calorimeter, and the distribution of energy
among the central cell and the surrounding ones depend on
the amount of signal deposited in the PS. This is shown in Figure~\ref{fig:psene}. By making use of the simulation, it was confirmed that this effect is mainly due to lateral leakage. This effect can be corrected in principle for the assessment of the response linearity. However, it is significantly harder to correctly account for it when assessing the resolution of the energy measurement. 

One particular run was taken with a 20-GeV positron beam with the PS detector out of the
beam line. After the requirement on the Cherenkov counters 
is applied, about $20000$ events are available for analysis, with 
negligible hadron contamination. 
Of these, about $4300$ are in the beam fiducial window defined above. This run (referred to as ``reference run'' in the following) was used to understand key features of the calorimeter response. 

\subsection{Positron energy measurement}

The selection of the positron component of the beam for the reference run was performed by requiring the signal in each of the Cherenkov threshold counters to be larger than five times the RMS of the corresponding pedestal. 

\begin{figure}[htb]
\begin{center}
\subfigure[]{
        \includegraphics[width=0.47\textwidth]{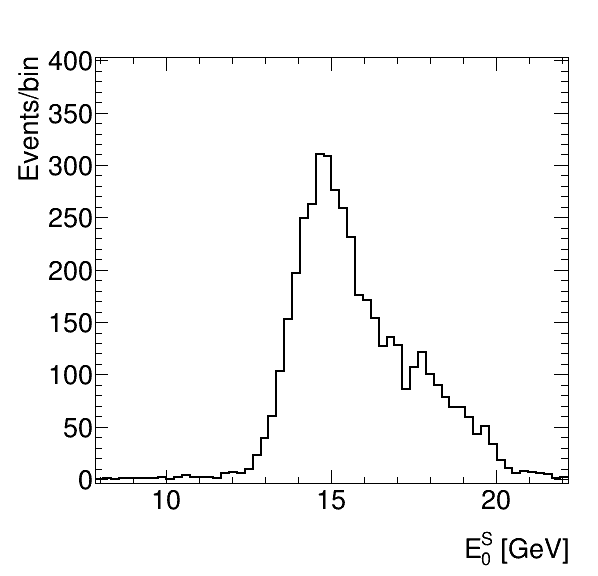}
        } \hfill
\subfigure[]{
        \includegraphics[width=0.47\textwidth]{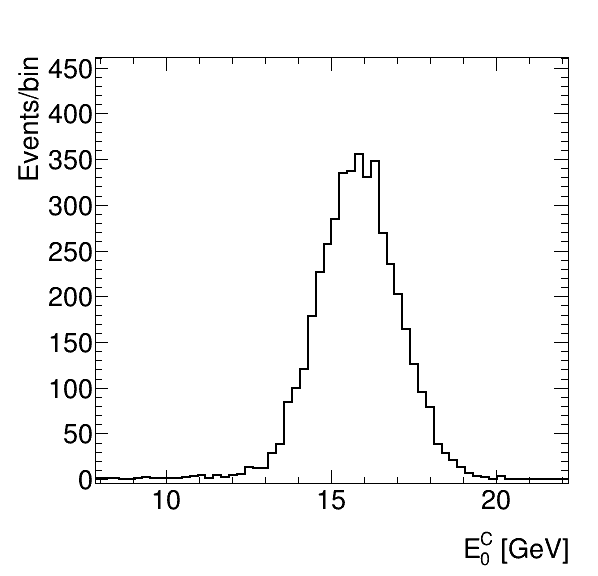}
        }
\caption{Distribution of the sums of the calibrated SiPM signals for the (a)
scintillating and (b) Cherenkov signals for the reference run.}
\label{fig:rawsipm}
\end{center}
\end{figure}

The distribution of the calibrated SiPM signals for the reference run is shown in Figure~\ref{fig:rawsipm} for the scintillation and Cherenkov signals.

\begin{figure}[htb]
\begin{center}
\subfigure[]{
        \includegraphics[width=0.47\textwidth]{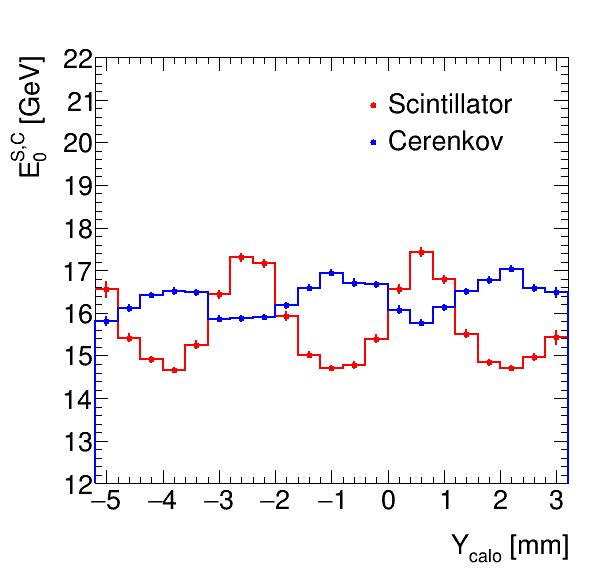}
        } \hfill
\subfigure[]{
        \includegraphics[width=0.47\textwidth]{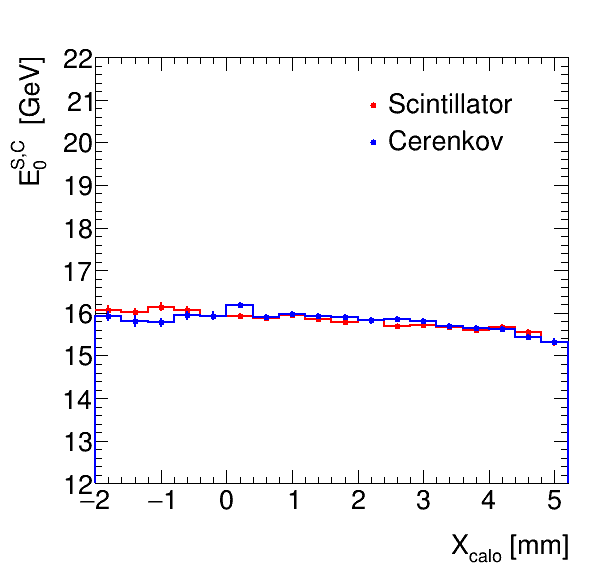}
        }
\caption{Average of the summed SiPM signals as a function of (a) \Ycalo and (b) \Xcalo for the reference run. The red points refer to the scintillation signal, while the blue ones refer to the Cherenkov.}
\label{fig:modulation}
\end{center}
\end{figure}

These distributions are far from being Gaussian. In order to understand this feature, the dependence of the energy measurement on the position of the barycentre of the shower was studied. It is shown in Figure~\ref{fig:modulation}.

A periodical modulation is observed in the \Ycalo coordinate. The period corresponds to the spacing between different types of fibres (Figure~\ref{fig:prototype}(c)). The amplitude of the oscillation is 
approximately $\pm10\%$ and $\pm5\%$ of the average signal for scintillator and
Cherenkov light, respectively. The phases of the oscillation of the scintillator and Cherenkov signals  
are opposite. No such modulation is observed in the $x$ coordinate.

The origin of the modulation is the same as discussed in Ref.~\cite{Akchurin:2005eu}: the modulation in $\Ycalo$ is due to the geometrical structure
of the calorimeter with alternate rows of scintillating and
Cherenkov fibres, combined with the absence of tilt between the fibres and the beam line in the $y$-direction. By summing the Cherenkov and scintillation signals the effect is mitigated, but
it does not disappear, as the observed oscillation amplitude is smaller
for the Cherenkov signal. This difference is ascribed to the fact that Cherenkov light 
has a well defined geometrical relationship with the direction
of the particle producing it, whereas scintillation light is produced nearly isotropic.

This result is correctly reproduced by the simulation: the amplitude and phase of the modulation have a strong dependence
on the impact angles of the beam on the calorimeter. The optimal agreement between the simulation and the data is obtained for an angle of 1.5$^{\circ}$ in $x$ and -0.4$^{\circ}$ in $y$, compatible with the setup described in Section~\ref{sec:det_descr}. 

\begin{figure}[htb]
\begin{center}
	\includegraphics[width=0.45\textwidth]{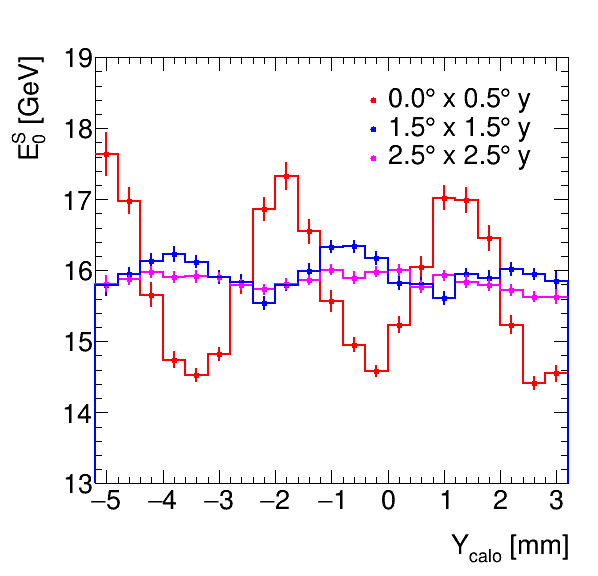}
\caption{Average of the energy in $\M{0}$ as a function of $\Ycalo$ for a simulated
20-GeV positron beam. The distribution is shown for three different configurations
of the beam angle into the calorimeter. The number in the legend correspond to the angle in the $x-z$  (indicated by $x$) and $y-z$ (indicated by $y$) planes. The same fiducial selection is applied as for data.}
\label{fig:angledep}
\end{center}
\end{figure}

The amplitude of the energy measurement modulation was studied as a function of the impact
angle of the beam onto the calorimeter. 
Figure~\ref{fig:angledep} shows the simulated
scintillation energy measurement as a function of the \Ycalo for three different beam angles in the $(y,z)$ plane. An angle of $2.5^{\circ}$ is sufficient to completely cancel the modulation.

Profiting from the high granularity of the optical readout in $\M{0}$, a correction procedure for the dependence of the calorimeter energy measurement on the particle impact point was developed by making use of the reference run. A useful variable to this aim is $\Rmax$, defined as the ratio of the energy deposited in $\M{0}$ in the row of scintillating fibres with the highest signal to the total scintillation signal in $\M{0}$.  

\begin{figure}[htb]
\begin{center}
\subfigure[]{
	\includegraphics[width=0.45\textwidth]{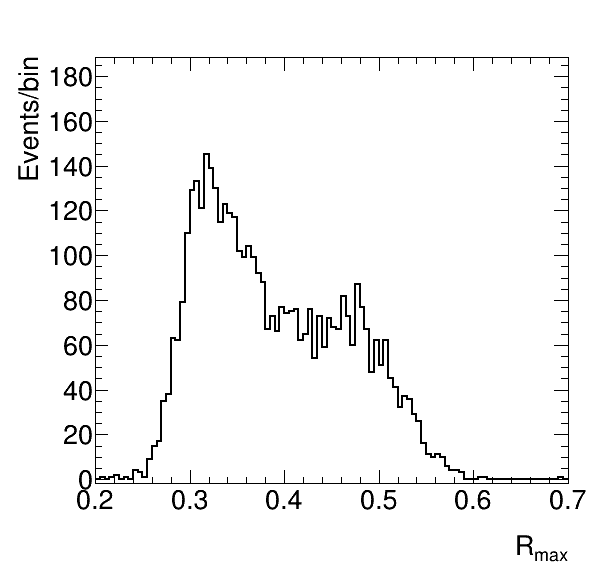}
 }\hfill
 \subfigure[]{
	\includegraphics[width=0.45\textwidth]{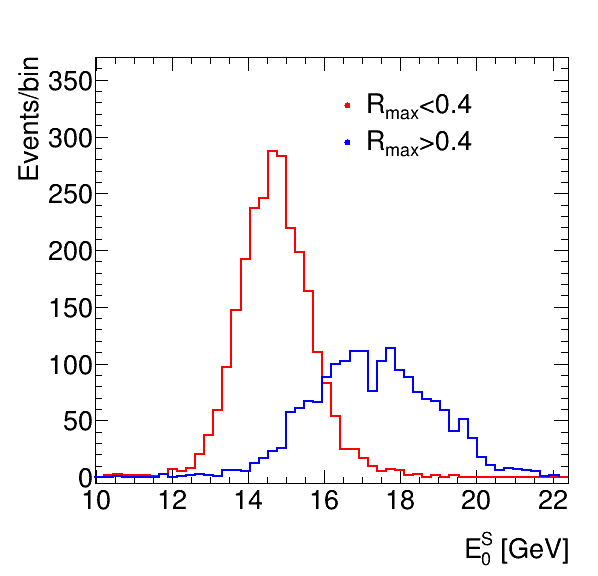}
 }
	\caption{(a) Distribution of $\Rmax$ for positrons in the reference run. (b) Distribution of the total scintillation energy signal  for events with $\Rmax<0.4$ (red) and 
	$\Rmax>0.4$ (blue).}
\label{fig:rmax}
\end{center}
\end{figure}

The distribution of $\Rmax$ for the reference run is 
shown in Figure~\ref{fig:rmax}(a). Figure~\ref{fig:rmax}(b) confirms that the events with high value of $\Rmax$ 
correspond to the configurations with high energy deposits in the
scintillation channel: a positron running approximately
parallel to the fibre direction in the vertical plane initiates a shower in a row corresponding to scintillating fibres, 
yielding a high scintillation response overall and high values of $\Rmax$. On the other hand, a shower initiated in Cherenkov fibre row, will share its energy between multiple scintillation rows, yielding a lower value for $\Rmax$.

\begin{figure}[htb]
\begin{center}
\subfigure[]{
	\includegraphics[width=0.45\textwidth]{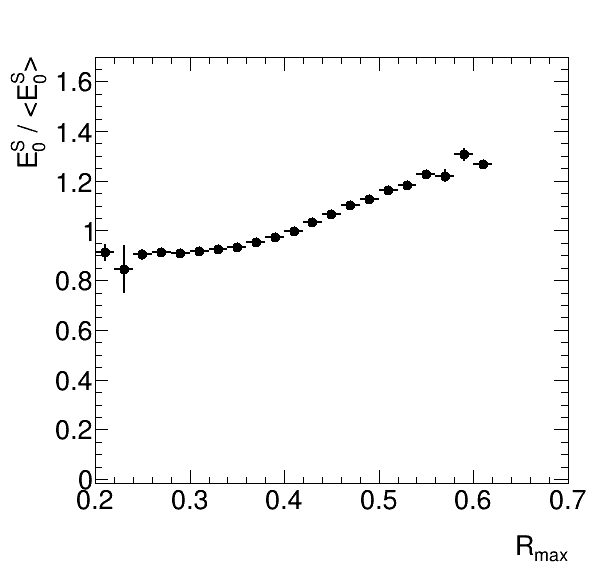}
} \hfill
\subfigure[]{
 \includegraphics[width=0.45\textwidth]{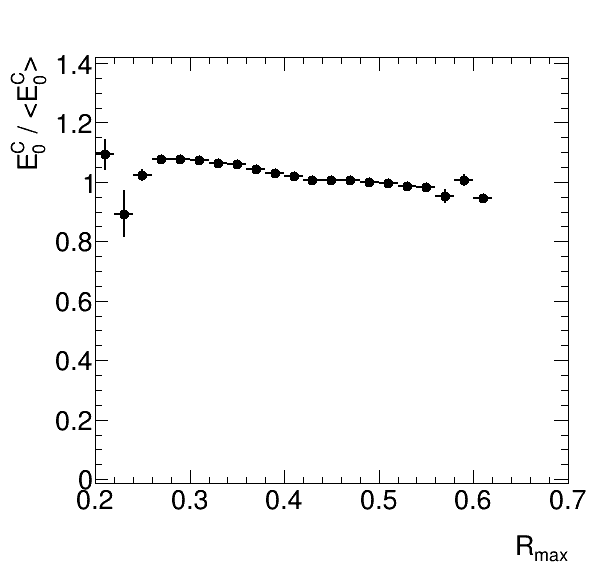}
 }
\caption{Average energy measured in the central cell normalised to the 
expected as a function of $\Rmax$ for the (a) scintillation  and (b) Cherenkov signal. 
}
\label{fig:rmaxdep}
\end{center}
\end{figure}

\begin{figure}[htb]
\begin{center}
\subfigure[]{
	\includegraphics[width=0.45\textwidth]{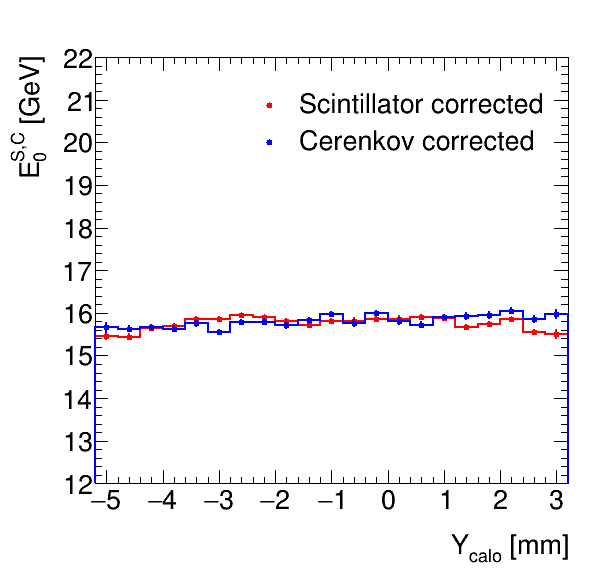}
 } \hfill
 \subfigure[]{
	\includegraphics[width=0.45\textwidth]{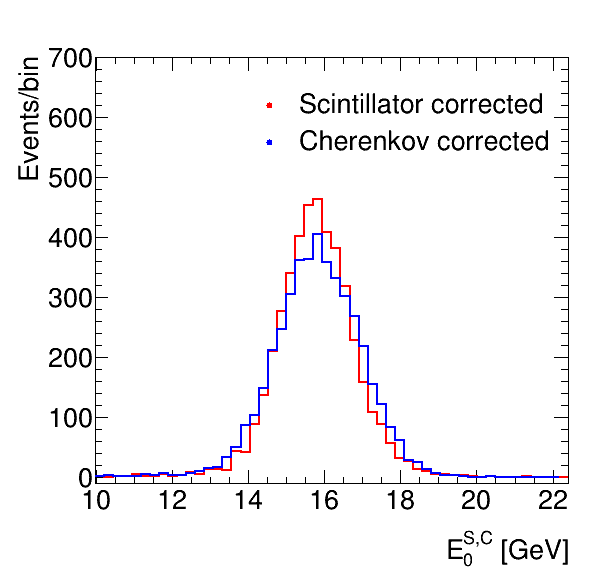}
 }
\caption{(a) Average of the summed SiPM signals in $\M{0}$ as a function of $\Ycalo$ for the reference run after
the correction procedure described in the text. The scintillation channels
are in red, the Cherenkov ones in blue. (b) Distribution of the sums of the calibrated SiPM signals for scintillating (red) and Cherenkov (blue) signals in $\M{0}$
after corrections for the reference run.
}
\label{fig:modcorr}
\end{center}
\end{figure}

The dependence on $\Rmax$ of the energy measured in $\M{0}$ is shown in Figure~\ref{fig:rmaxdep}
for the scintillation and the Cherenkov signals. Such dependence is parametrised with a second-degree polynomial,
and the resulting correction factor is applied to the measured energy.
The result of the procedure is shown in Figure~\ref{fig:modcorr} (a): the position dependence of the measurement is 
almost completely removed. The distribution for the measured energies in $\M{0}$ after this correction is
shown in Figure~\ref{fig:modcorr} (b), for both the scintillation and Cherenkov signals. 
The correction recovers a nearly Gaussian response for both the scintillation and Cherenkov signals.

For the following results, the procedure just described is applied to the data and, where relevant, to the simulation.

\subsection{Determination of the calorimeter response to positrons}

Pure samples of positrons for all available beam energies between 10 and 100 GeV can be selected by applying the fiducial selection described in Section~\ref{sec:elec_selection}. For beam energies up to $\Ebeam = 30\ \mathrm{GeV}$, the positron component of the beam was selected by requiring the signal in each of the Cherenkov threshold counters to be larger than five times the RMS of the corresponding pedestal. For energies $\Ebeam > 30\ \mathrm{GeV}$, the positron selection was done by requiring a signal in the PS larger or equal to three MIPs. The dependence of the 
average energy deposit on the number of MIPs deposited in the
preshower shown in Figure~\ref{fig:psene} is corrected for by adding to the total calorimeter energy a linear function of the PS signal, determined with a linear fit. This correction is indicated in the following as $\delta_{\mathrm{PS}}$. The correction is about $80\ \mathrm{MeV/MIP}$. The variable $E_{\mathrm{meas}}$ is therefore defined as 

\begin{align*}
    E_{\mathrm{meas}} = \delta_{\mathrm{PS}} + \frac{\sum_i E^{S}_i + \sum_i E^{C}_i}{2}
\end{align*}

\begin{figure}[htb]
\begin{center}
	\includegraphics[width=0.6\textwidth]{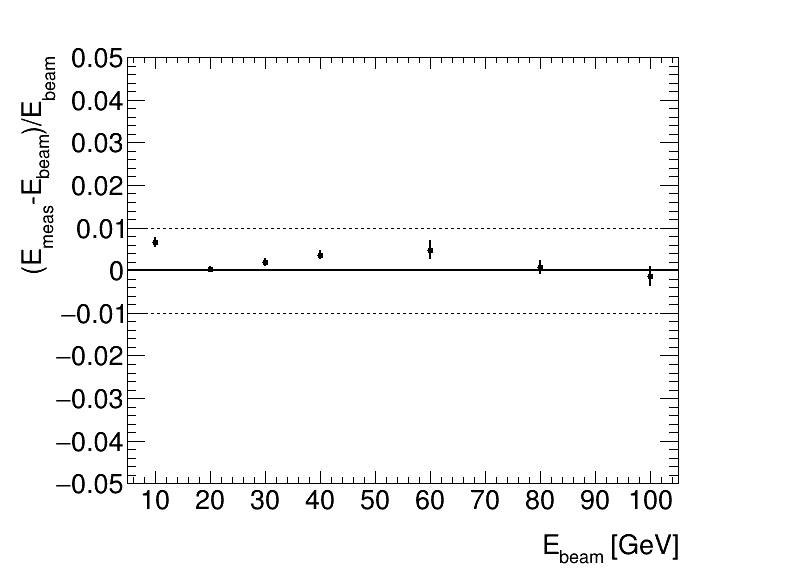}
\caption{Dependency of the calorimeter response on the beam energy. 
}
\label{fig:calolin}
\end{center}
\end{figure}

The linearity of the calorimeter response is shown in Fig~\ref{fig:calolin}. The linearity of the response was better than 1\% over the full explored range.

\subsection{Resolution measurement}

The impact of the PS selection on the resolution of the energy measurement is very strong, and it leads to calorimeter performance which are significantly worse than those that one would have obtained in absence of PS-induced lateral leakage. For this reason, a different strategy was adopted: for the determination of the energy resolution, only datasets corresponding to beam energies where the positron selection can reliably be done only by using the Cherenkov counters were used. This limits the available energies to 10, 20 and 30~GeV. On top of the Cherenkov counter selection, an additional requirement to accept only events leaving a signal equivalent to one MIP in the PS was imposed. The obtained resolution was compared to that of the simulation. Finally, the simulation is used to extrapolate at higher energies.

\begin{figure}[htb]
\begin{center}
\subfigure[]{
	\includegraphics[width=0.45\textwidth]{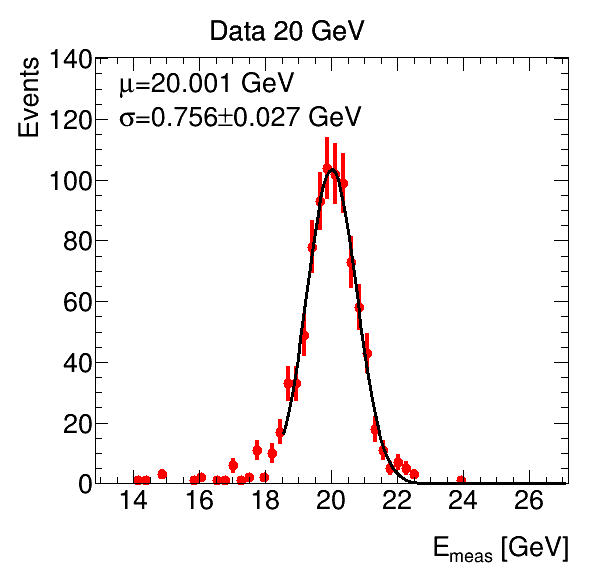}
 } \hfill
 \subfigure[]{
	\includegraphics[width=0.45\textwidth]{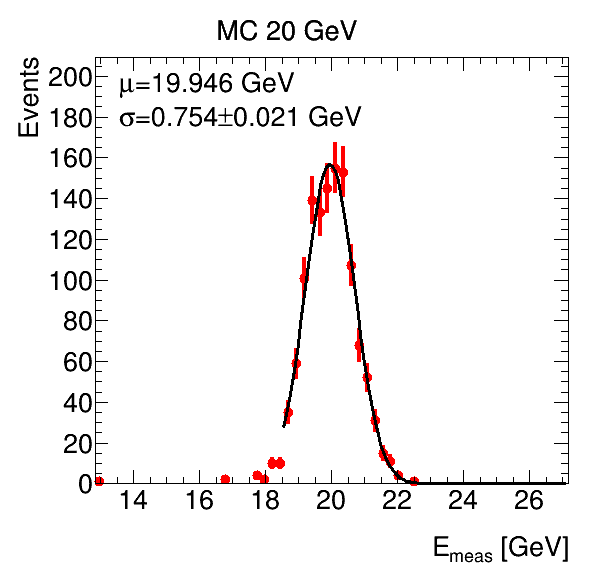}
 }
\caption{Distribution of the energy measured after all corrections for a 20-GeV 
positron beam for (a) data and (b) the simulation.}
\label{fig:peak20}
\end{center}
\end{figure}

The energy measurement of the calorimeter distribution of the energy measurement response obtained for a 20-GeV run with the PS is shown in Figure~\ref{fig:peak20} for data and simulation.

\begin{figure}[htb]
\begin{center}
	\includegraphics[width=0.7\textwidth]{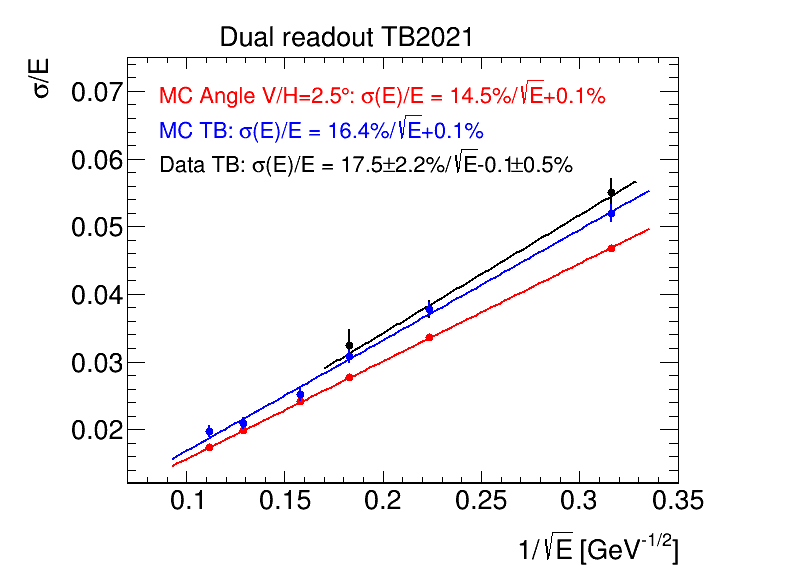}
\caption{Measured resolution of the calorimeter prototype for positrons
as a function of the reciprocal of the square root of the beam energy.
The black points represent the data, the blue ones correspond to the simulation  reproducing
the test beam geometry including the PS, the red ones correspond to a simulation  with
	the beam impinging on the calorimeter with an angle of 2.5$^{\circ}$ 
both in the horizontal and in the vertical direction and no PS on the beam line.}
\label{fig:resfin}
\end{center}
\end{figure}

The calorimeter resolution is evaluated as the ratio of $\sigma/E$ of the RMS width to the mean value of a Gaussian fit to the energy measurement distribution for data and simulation.  The results are summarised in Figure~\ref{fig:resfin}. For the three available 
experimental points the measured resolution is compatible with the 
simulated one. For comparison, the resolution of the 
same module  in a setup with no preshower in the beam is also calculated,
with an angle of impact of 2.5$^{\circ}$ in both the horizontal and 
vertical direction. 

The resolution for the simulation reproducing the layout of the test beam 
has a stochastic term of 16.4\% and a constant term of 0.1\%. 
The three available data points yield  resolutions compatible
with the simulation. Because of the small constant term, fits with a linear or quadrature sum of the stochastic and constant term yield similar results. 

The simulation with a 2.5$^{\circ}$ angle yields a stochastic term 
of 14.5\% which we consider representative of the electromagnetic
resolution of the tested geometry in a real experiment.

\subsection{Positron Shower Profile}

Thanks to the single fibre readout of the central module and the small distance of 2~mm between the individual fibres, a high-granularity measurement of the lateral development of the electromagnetic shower could be compared to the predictions of the Geant4 simulation. The measurement was performed independently for the scintillation and Cherenkov signal. Due to the high statistics needed to extract shower profiles, only the reference run was used for this measurement. Similarly to the event selection described before, positrons are selected by the usage of the Cherenkov counters. Moreover only events with particles hitting the centre of the calorimeter front face within a radius of $5$ mm are kept for analysis. About $4\times10^3$ events from the reference run are selected for analysis and are compared to $50\times10^3$ simulated events\footnote{For this simulation the patch03 of the \Geant 10.7 release was used.}. Only the SiPMs' signals were considered, therefore the profiles in the following refer to the shower's core with an energy containment of $72\%$. 
 Since one run only was used, the SiPM signals were calibrated using dedicated calibration constants obtained from events of the reference run only. \\ 

\begin{figure}[htb]
\begin{center}
	\includegraphics[width=0.7\textwidth]{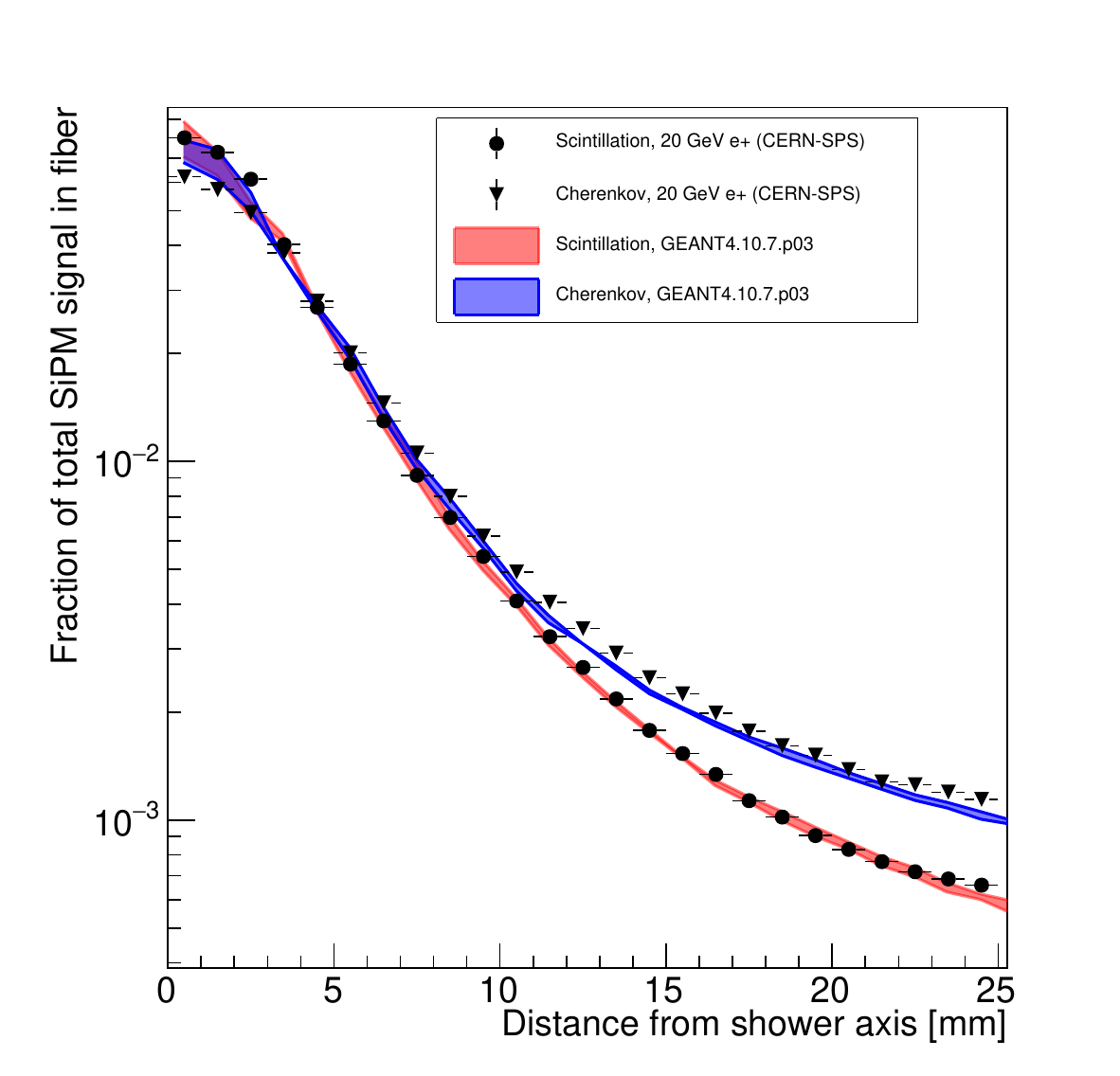}
\caption{Lateral shower profile of positrons with an energy of 20 GeV. The black circles (triangles) refer to the scintillation (Cherenkov) signal in $\M{0}$. The red and blue bands correspond to the simulation predictions for the scintillation and Cherenkov signal,  respectively.}
\label{fig:lateral_profile}
\end{center}
\end{figure}

The {\em lateral shower profile} was defined in Ref.~\cite{Antonello:2018sna} as the fraction of energy deposited in a single fibre at a distance from the shower axis, defined as

\begin{align*}
    r = \sqrt{\left(x - \Xcalo\right)^2 + \left(y - \Ycalo\right)^2} 
\end{align*}

\noindent for a fibre whose position is $\left(x,y\right)$. The measurement is compared to the simulation predictions in Figure~\ref{fig:lateral_profile}. It is remarkable that the average signal carried by a fibre drops by two orders of magnitude over a radial distance of only $25$ mm. For the simulated results, an error of $\pm 0.4^{\circ}$ in the orientation of the calorimeter with respect to beam axis is considered: the horizontal angle was assumed to be within $1.0^{\circ}$ and $1.8^{\circ}$ while the vertical angle was not changed with respect to previous simulations. This assumption is compatible with the experimental setup described Section~\ref{sec:elec_selection} and leads to the colored error bands. The good agreement between the simulation predictions and the data over the entire range probed gives confidence on the quality of the calorimeter simulation.  

\begin{figure}[htb]
\begin{center}
\subfigure[]{
	\includegraphics[width=0.48\textwidth]{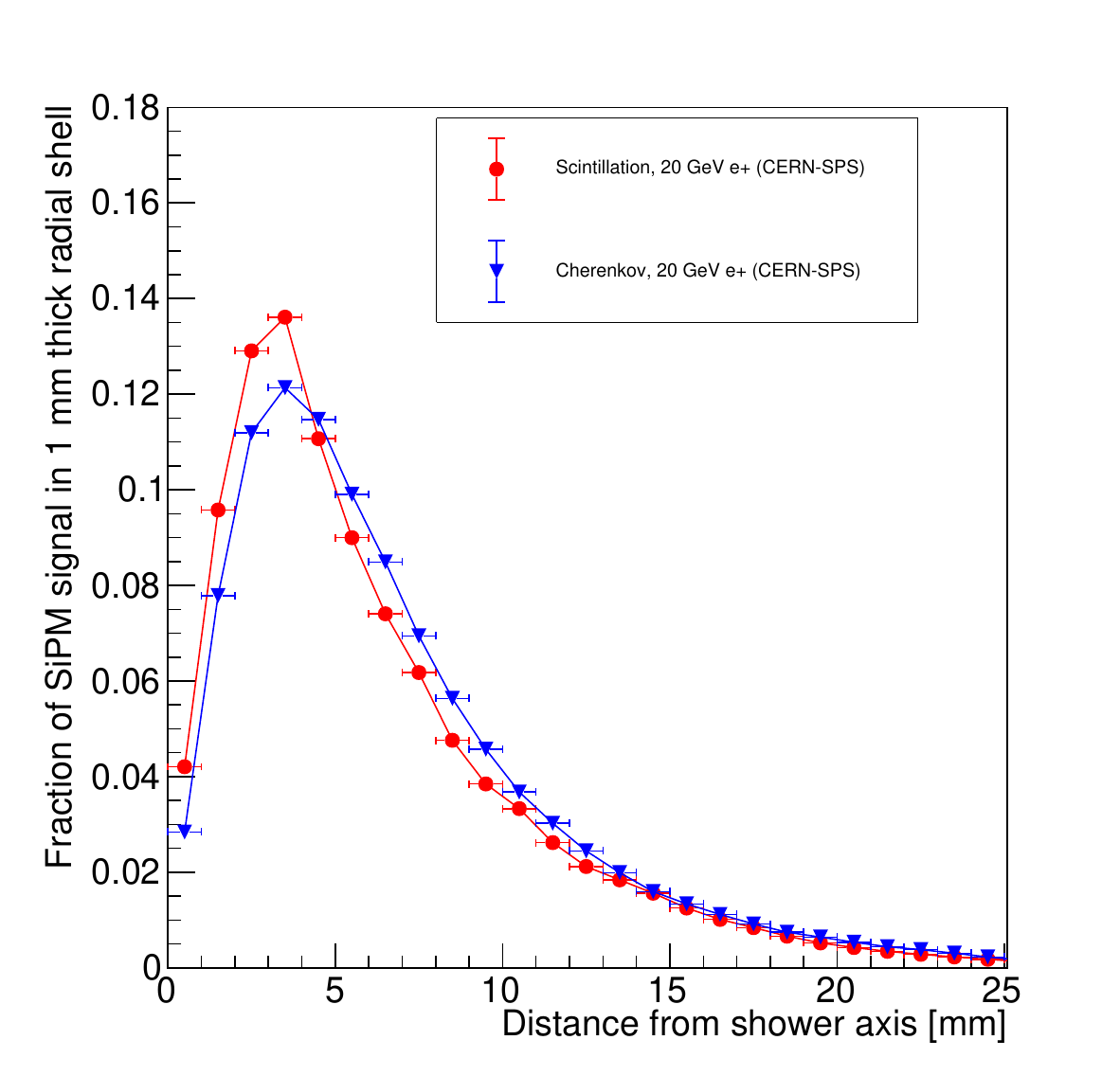}\label{fig:radial_profile_a}
 } \hfill
 \subfigure[]{
        \includegraphics[width=0.48\textwidth]{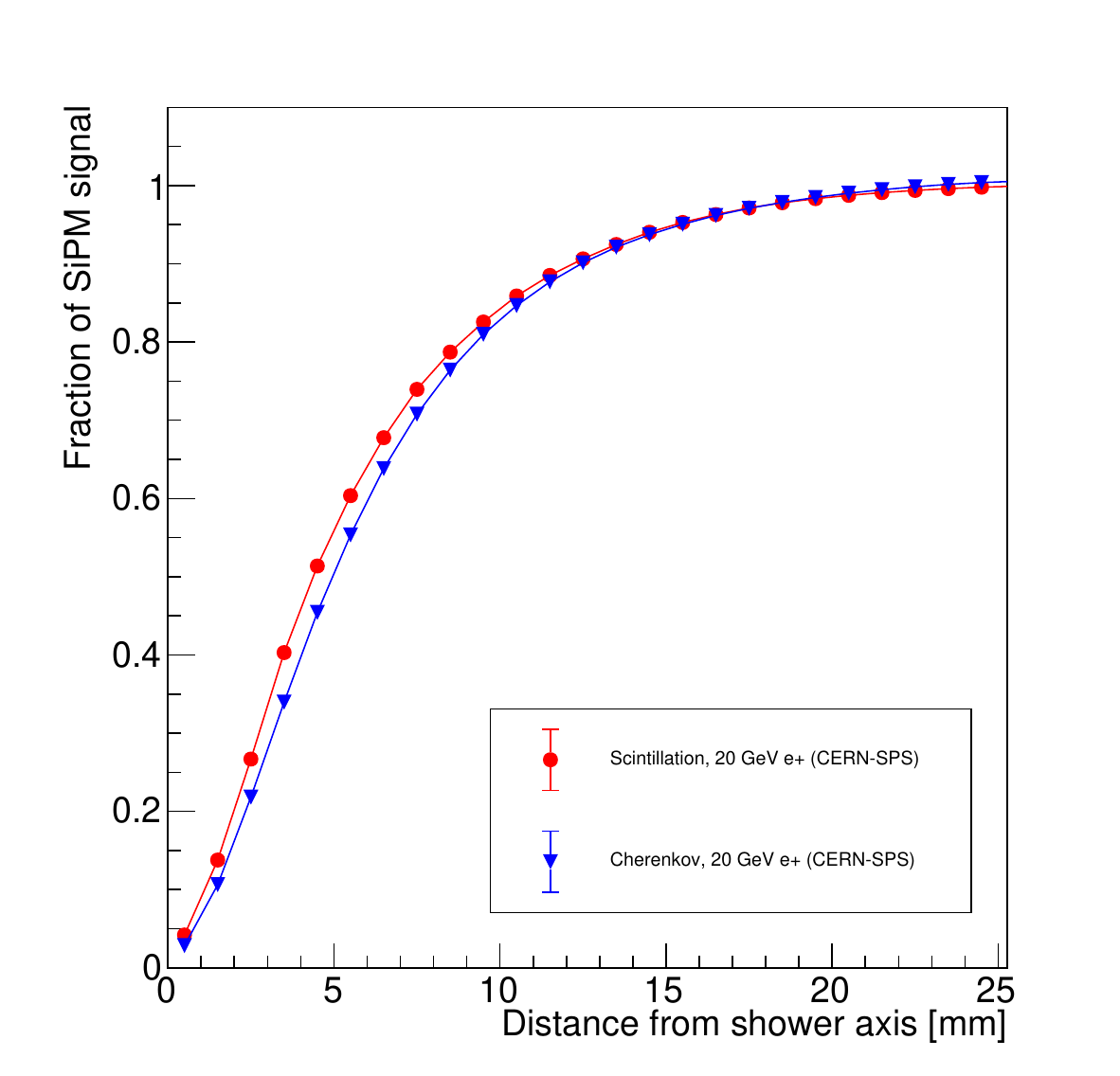}\label{fig:radial_profile_b}
        }
\caption{(a) Radial and (b) cumulative shower profiles of positrons with an energy of 20 GeV. The red circles (blue triangles) refer to the scintillation (Cherenkov) signal in $\M{0}$.}
\label{fig:radial_profile}
\end{center}
\end{figure}

The shower profile as seen by the Cherenkov signal is wider than the one measured with the scintillation light. This observation confirms those of Refs.~\cite{Akchurin:2005rs} (for a different calorimeter setup) and \cite{Antonello:2018sna}. This is understood to be due to the fact that the early components of electromagnetic showers are collimated with the incoming positron direction and the Cherenkov emitted light falls outside the numerical aperture of the fibre.\\
Figure~\ref{fig:radial_profile}, derived from the same experimental data, provides different views on the same measurement. Ref.~\cite{Antonello:2018sna} defined the {\em radial shower profile} by the fraction of energy deposited in a ring around the shower axis, as a function of the distance from the shower axis. To this extent, the energy in all fibres in a $r$ bin was summed and divided by the total shower energy in $\M{0}$, and its average plotted against the distance from the shower axis. The result is shown in Figure~\ref{fig:radial_profile_a}. The corresponding cumulative distribution (obtained by summing the contributions to the shower up to a given distance from the shower axis) is shown in Figure~\ref{fig:radial_profile_b}. Both figures illustrate the effect that the shower appears wider when measured using the Cherenkov contribution.

\section{Conclusions}
\label{sec:conclusions}
A dual-readout sampling calorimeter prototype using brass capillary tubes as absorber and optical fibres as active medium was tested using beams of particles at DESY and at the H8 beam line at CERN. The dual readout was realised by making use of two different types of fibres: doped scintillating Saint-Gobain BCF-10 fibres, and clear ``Cherenkov'' Mitsubishi SK40 fibres. The prototype (with a total size of about $10\times10\times 100\ \mathrm{cm^3}$) was composed by nine modules. For the central module, the individual fibres were read out by means of Hamamatsu S14160-1315 PS SiPMs, while for the surrounding eight modules the two sets of fibres were bundled together and read out by Hamamatsu R8900 PMTs. 

After having calibrated the detector by making use of the SiPM multiphoton spectrum and of beams of positrons, a procedure was developed to correct the detector response for the dependency of the energy measurement on the particle impact point on the calorimeter front face. Then, the detector response was studied using beams of positrons with energies between 10 and 100~GeV: the linearity of the energy measurement was found to be within 1\%, the measured response was found to be in agreement with that predicted by a dedicated Geant4 simulation. The energy resolution was found to have a significant dependence on the impact angle of the beam on the calorimeter front face for angles between $\theta = 0^{\circ}$ and $\theta = 2.5^{\circ}$. The energy resolution on data was evaluated by making use of beams of positrons with $\Ebeam = 10, 20\ \mathrm {and}\ 30\ \mathrm{GeV}$, and determined to be 

\begin{align*}
    \frac{\sigma}{E} = \frac{\left(17.5\pm 2.2\right)\%}{\sqrt{E}} + \left(0.1\pm 0.5\right)\%,
\end{align*}

\noindent in agreement with that predicted by the simulation for the same beam impact angle. The extrapolation performed with the simulation up to impact angles $\theta = 2.5^{\circ}$ showed that the optimal resolution of the prototype is expected to be

\begin{align*}
    \frac{\sigma}{E} = \frac{14.5\%}{\sqrt{E}} + 0.1\%.
\end{align*}

Finally, a precise measurement of the electromagnetic shower profile of 20-GeV positrons was compared with the predictions of the simulation, finding good agreement. The shower development as seen by the scintillation signal is narrower than that seen by the Cherenkov one, as a consequence of the loss of Cherenkov light from particles in the early phases of  the shower development. 

The electromagnetic performance of the dual-readout sampling calorimeter prototype is overall found to be satisfactory and in line with the expectations, making this mechanical and readout solution a promising one for future developments.

\bibliography{DRtubes}

\providecommand{\href}[2]{#2}\begingroup\raggedright\begin{thebibliography}{10}

\bibitem{FCC:2018evy}
{\scshape FCC} collaboration, \emph{{FCC-ee: The Lepton Collider}: {Future
  Circular Collider Conceptual Design Report Volume 2}},
  \href{https://doi.org/10.1140/epjst/e2019-900045-4}{\emph{Eur. Phys. J. ST}
  {\bfseries 228} (2019) 261}.

\bibitem{CEPCStudyGroup:2018ghi}
{\scshape CEPC Study Group} collaboration, \emph{{CEPC Conceptual Design
  Report: Volume 2 - Physics \& Detector}},
  \href{https://arxiv.org/abs/1811.10545}{{\ttfamily 1811.10545}}.

\bibitem{Lee:2017xss}
S.~Lee, M.~Livan and R.~Wigmans, \emph{{Dual-Readout Calorimetry}},
  \href{https://doi.org/10.1103/RevModPhys.90.025002}{\emph{Rev. Mod. Phys.}
  {\bfseries 90} (2018) 025002}
  [\href{https://arxiv.org/abs/1712.05494}{{\ttfamily 1712.05494}}].

\bibitem{Akchurin:2005eu}
N.~Akchurin, K.~Carrell, H.~Kim, R.~Thomas, R.~Wigmans, J.~Hauptman et~al.,
  \emph{{Electron detection with a dual-readout calorimeter}},
  \href{https://doi.org/10.1016/j.nima.2004.06.178}{\emph{Nucl. Instrum. Meth.
  A} {\bfseries 536} (2005) 29}.

\bibitem{Akchurin:2005an}
N.~Akchurin, K.~Carrell, J.~Hauptman, H.~Kim, H.P.~Paar, A.~Penzo et~al.,
  \emph{{Hadron and jet detection with a dual-readout calorimeter}},
  \href{https://doi.org/10.1016/j.nima.2004.07.285}{\emph{Nucl. Instrum. Meth.
  A} {\bfseries 537} (2005) 537}.

\bibitem{Akchurin:2005rs}
N.~Akchurin, K.~Carrell, H.~Kim, R.~Thomas, R.~Wigmans, J.~Hauptman et~al.,
  \emph{{Comparison of high-energy electromagnetic shower profiles measured
  with scintillation and Cherenkov light}},
  \href{https://doi.org/10.1016/j.nima.2005.03.169}{\emph{Nucl. Instrum. Meth.
  A} {\bfseries 548} (2005) 336}.

\bibitem{Akchurin:2013yaa}
N.~Akchurin et~al., \emph{{Dual-readout Calorimetry}},
  \href{https://arxiv.org/abs/1307.5538}{{\ttfamily 1307.5538}}.

\bibitem{Lee:2017shn}
S.~Lee et~al., \emph{{Hadron detection with a dual-readout fiber calorimeter}},
  \href{https://doi.org/10.1016/j.nima.2017.05.025}{\emph{Nucl. Instrum. Meth.
  A} {\bfseries 866} (2017) 76}
  [\href{https://arxiv.org/abs/1703.09120}{{\ttfamily 1703.09120}}].

\bibitem{Akchurin:2014aoa}
N.~Akchurin et~al., \emph{{The electromagnetic performance of the RD52 fiber
  calorimeter}}, \href{https://doi.org/10.1016/j.nima.2013.09.033}{\emph{Nucl.
  Instrum. Meth. A} {\bfseries 735} (2014) 130}.

\bibitem{Antonello:2018sna}
M.~Antonello et~al., \emph{{Tests of a dual-readout fiber calorimeter with SiPM
  light sensors}},
  \href{https://doi.org/10.1016/j.nima.2018.05.016}{\emph{Nucl. Instrum. Meth.
  A} {\bfseries 899} (2018) 52}
  [\href{https://arxiv.org/abs/1805.03251}{{\ttfamily 1805.03251}}].

\bibitem{lorenzoPhD}
L.Pezzotti, \emph{Particle Detectors R\&D: Dual-Readout Calorimetry for Future
  Colliders and MicroMegas Chambers for the ATLAS New Small Wheel Upgrade},
  Ph.D. thesis, {Universit\'a degli Studi di Pavia}, 2021.

\bibitem{Pezzotti:2022ndj}
I.~Pezzotti et~al., \emph{{Dual-Readout Calorimetry for Future Experiments
  Probing Fundamental Physics}},
  \href{https://arxiv.org/abs/2203.04312}{{\ttfamily 2203.04312}}.

\bibitem{Gaudio:2022jve}
G.~Gaudio, \emph{{The IDEA detector concept for FCCee}},
  \href{https://doi.org/10.22323/1.414.0337}{\emph{PoS} {\bfseries ICHEP2022}
  (2022) 337}.

\bibitem{Lucchini:2020bac}
M.T.~Lucchini, W.~Chung, S.C.~Eno, Y.~Lai, L.~Lucchini, M.-T.~Nguyen et~al.,
  \emph{{New perspectives on segmented crystal calorimeters for future
  colliders}},
  \href{https://doi.org/10.1088/1748-0221/15/11/P11005}{\emph{JINST} {\bfseries
  15} (2020) P11005} [\href{https://arxiv.org/abs/2008.00338}{{\ttfamily
  2008.00338}}].

\bibitem{Lucchini:2022vss}
M.T.~Lucchini, L.~Pezzotti, G.~Polesello and C.G.~Tully, \emph{{Particle flow
  with a hybrid segmented crystal and fiber dual-readout calorimeter}},
  \href{https://doi.org/10.1088/1748-0221/17/06/P06008}{\emph{JINST} {\bfseries
  17} (2022) P06008} [\href{https://arxiv.org/abs/2202.01474}{{\ttfamily
  2202.01474}}].

\bibitem{Karadzhinova-Ferrer:2022paf}
A.~Karadzhinova-Ferrer et~al., \emph{{Novel prototype tower structure for the
  dual-readout fiber calorimeter}},
  \href{https://doi.org/10.1088/1748-0221/17/09/T09007}{\emph{JINST} {\bfseries
  17} (2022) T09007}.

\bibitem{BC10}
{Saint Gobain BC-10}.
  \url{https://www.crystals.saint-gobain.com/radiation-detection-scintillators/fibers},
  accessed on 14 March 2023.

\bibitem{SK40}
{Mitsubishi ESKA SK40}.
  \url{https://www.pofeska.com/pofeskae/download/pdf/f/SK40.pdf}, accessed on
  14 March 2023.

\bibitem{R8900}
{HAMAMATSU R8900}.
  \url{https://www.hamamatsu.com/content/dam/hamamatsu-photonics/sites/documents/99_SALES_LIBRARY/etd/R8900(U)-00-C12_TPMH1299E.pdf},
  accessed on 14 March 2023.

\bibitem{S14160}
{HAMAMATSU S14160-1315PS}.
  \url{https://www.hamamatsu.com/jp/en/product/optical-sensors/mppc/mppc_mppc-array/S14160-1315PS.html},
  accessed on 16 March 2023.

\bibitem{FERS:CAEN}
{FERS A5202, CAEN S.p.A., FERS A5202}.
  \url{https://www.caen.it/products/a5202/}, accessed on 18 January 2023.

\bibitem{CITIROC}
Citiroc-1A. \url{https://www.weeroc.com/products/sipm-read-out/citiroc-1a},
  accessed on 18 January 2023.

\bibitem{Diener:2018qap}
R.~Diener et~al., \emph{{The DESY II Test Beam Facility}},
  \href{https://doi.org/10.1016/j.nima.2018.11.133}{\emph{Nucl. Instrum. Meth.
  A} {\bfseries 922} (2019) 265}
  [\href{https://arxiv.org/abs/1807.09328}{{\ttfamily 1807.09328}}].

\bibitem{Dannheim:2013iea}
{\scshape Linear Collider CERN Detector} collaboration, \emph{{Particle
  Identification with Cherenkov detectors in the 2011 CALICE Tungsten Analog
  Hadronic Calorimeter Test Beam at the CERN SPS}}, {\emph{LCD-Note-2013-006,
  AIDA-NOTE-2015-012} (2013) }.

\bibitem{Agostinelli:2002hh}
{GEANT4 Collaboration}, S.~Agostinelli et~al., \emph{{\textsc{Geant4} -- a
  simulation toolkit}},
  \href{https://doi.org/10.1016/S0168-9002(03)01368-8}{\emph{Nucl. Instrum.
  Meth. A} {\bfseries 506} (2003) 250}.

\bibitem{Giaz:2023qsm}
A.~Giaz et~al., \emph{{Test beam results of the fiber-sampling dual-readout
  calorimeter}}, \href{https://doi.org/10.1016/j.nima.2022.167964}{\emph{Nucl.
  Instrum. Meth. A} {\bfseries 1048} (2023) 167964}.

\end{thebibliography}\endgroup
\bibliographystyle{JHEP}

\end{document}